\documentclass{aa}

\usepackage{graphicx}
\usepackage{xcolor}
\usepackage{txfonts}
\usepackage{hyperref}
\usepackage{bm}
\usepackage{soul}
\usepackage{orcidlink}
\usepackage{tabularx}
    \newcolumntype{L}{>{\raggedright\arraybackslash}X}

\newcommand{\legolas}{\textsf{Legolas}}
\newcommand{\amrvac}{\textsf{MPI-AMRVAC}}
\newcommand{\pluto}{\textsf{PLUTO}}
\newcommand{\muram}{\textsf{MURaM}}
\newcommand{\lare}{\textsf{Lare}}
\newcommand{\athena}{\textsf{Athena++}}

\newcommand{\bfb}{\bm{B}}

\newcommand{\bfv}{\bm{v}}

\newcommand{\bfk}{\bm{k}}
\newcommand{\bfx}{\bm{x}}

\newcommand{\ex}{\hat{\bm{e}}_x}
\newcommand{\ey}{\hat{\bm{e}}_y}
\newcommand{\ez}{\hat{\bm{e}}_z}
\newcommand{\im}{\mathrm{i}}

\newcommand{\dA}{\,\mathrm{d}A}

\allowdisplaybreaks

\begin{document} 

   \title{Eigenmode initialisation of 2D (magneto)hydrodynamic simulations}

   \subtitle{}

   \author{J. De Jonghe\inst{1,2} \orcidlink{0000-0003-2443-3903} and A. J. B. Russell\inst{2} \orcidlink{0000-0001-5690-2351}}

   \institute{Centre for mathematical Plasma Astrophysics, KU Leuven, Celestijnenlaan 200B, Leuven, B-3001, Belgium
   \and School of Mathematics and Statistics, University of St Andrews, North Haugh, St Andrews, KY16 9SS, United Kingdom\\
              \email{jordi.dejonghe@kuleuven.be}}

   \date{Received 23 October 2024; accepted 22 February 2026}
 
  \abstract
   {}
   {The early evolution of unstable hydrodynamic and magnetohydrodynamic equilibria is often governed by a few dominant linear eigenmodes. We investigate whether initialising a simulation with a superposition of linear eigenmodes that contains the most unstable mode saves computation time, and how the selection of the included modes affects the non-linear evolution. Three 2D setups are considered and each is representative of a distinct non-linear behaviour: a flow-sheared fluid interface (density gradient) that is unstable to the Kelvin-Helmholtz instability, to study the effect of the initial perturbation on developing turbulence; a tearing-unstable Harris current sheet, to investigate how the initial perturbation affects secondary instabilities; and a magnetised flow-sheared plasma interface that is linearly unstable, but where magnetic tension suppresses the instability again in the non-linear phase.}
   {Using the non-linear (magneto)hydrodynamic simulation code \amrvac{}, the evolutions of a flow-sheared fluid interface, a Harris current sheet, and a flow-sheared plasma interface were simulated for various initial perturbations. The novel initial perturbations were linear eigenmodes of the equilibrium, or superpositions thereof, and calculated with the \legolas{} code. We benchmarked to initialisation with velocity noise and, in the case of the Harris sheet, initialisation with an analytic magnetic field perturbation.}
   {By perturbing an unstable equilibrium with a superposition of linear eigenmodes that contains the most unstable mode, significantly less computation time is spent on the linear stage of the evolution compared to traditional perturbation methods. In the best case, the simulation time needed to reach the non-linear stage is reduced by an order of magnitude. The inclusion or omission of certain modes in the initial perturbation is observed to affect the non-linear evolution to various degrees.}
   {The perturbation of equilibria with a superposition of linear eigenmodes that contains the most unstable mode allows simulations to reach a late-evolution stage faster, thus saving computation time. Additional eigenmodes can be included alongside the fastest growing mode to obtain further benefits, for example, to accelerate symmetry breaking in the non-linear stage, or to delay their effect on the non-linear evolution. Coupling spectroscopic codes with (magneto)hydrodynamic codes therefore offers significant advantages to the astrophysics community.}


   \maketitle

\section{Introduction}
In the last few decades, a lot of theoretical research in solar physics and astrophysics has heavily relied on hydrodynamic (HD) and  magnetohydrodynamic (MHD) simulations. This is clearly shown by the plethora of codes that are developed for the single purpose of resolving the dynamics of astrophysical plasma structures, such as the general-purpose MHD codes \athena{} \citep{Stone2020}, \lare{} \citep{Arber2001}, \amrvac{} \citep{Porth2014,Xia2018,Keppens2023}, and \pluto{} \citep{Mignone2007,Mignone2012}, or the more problem-specific \muram{} \citep{Vogler2005} code for the solar atmosphere.

Most astrophysicists involved in simulation and modelling are engaged with setting up existing codes to address specific problems, which \citet{HarlowFromm1965} called the `third step' of computer experiments. At this stage, scientists with detailed knowledge of the problem and science goals must decide which physics to include, as well as the parameter values, grid structure, resolution, boundary conditions, and initialisation. Every one of these decisions impacts the success of the numerical experiment, spanning from how efficiently computational resources are used, to whether or not the project achieves its objectives.

In our experience, the importance of selecting the minimum necessary physics and of using appropriate boundary conditions is generally considered to be self-evident, and the influence of parameter values and grid design is widely discussed. However, whilst it similarly contributes to a simulation's success, the topic of how best to initialise astrophysical simulations has received comparatively little attention, creating an opportunity to significantly improve the current practice.

In this paper, we focus on HD and MHD simulations that begin with an unstable equilibrium. Astrophysical use cases include studying the linear and non-linear development of instabilities, as well as later-stage phenomena such as turbulence, mixing, the production of seed regions for star formation, and the onset of fast magnetic reconnection. Since instabilities take a long time to grow from round-off error alone, it is standard practice to add perturbations to the initial state. This paper highlights that although the choice of the initial perturbation might at first seem to be a minor detail, it can dramatically impact the efficiency and effectiveness of simulation studies. It also presents examples that show the effectiveness of initialising with a superposition of linear eigenmodes, which reduces the simulation time needed to reach the non-linear stage by up to an order of magnitude.

Sub-optimally initialised simulations often achieve their science objectives less well than they might have. Runtime savings from better initialisation can be spent elsewhere, for example, increasing resolution to access more realistic parameter values or to resolve a larger turbulence inertial range, or to follow the non-linear dynamics for longer. In simulations that spend a long time in the linear growth phase, physical or numerical diffusion can alter the equilibrium whilst waiting for the instability to grow, such that the unstable modes that are eventually detected do not correspond to the unstable modes of the initial system. Another issue is that oscillations excited by an inexact perturbation may impact the later dynamics in subtle ways, complicating interpretation and limiting reproducibility.

In the most challenging applications, spending unnecessary computation time in the linear growth phase can prevent research projects from achieving their goals entirely. This cannot always be solved with greater computing power, because as with physical experiments, it can sometimes be desirable to perform numerical experiments that have an intrinsic maximum code time. Hence, the phenomena of interest must be produced as early as possible, requiring careful attention to the initialisation. An example of this arises in one of the authors' own research areas, studying fast magnetic reconnection between two merging flux ropes \citep[e.g.][]{Beg2022}, which requires that instabilities trigger fast reconnection as early as possible before the flux ropes have fully merged, or in some cases before other instabilities that operate on a long timescale have grown significantly.

Even in cases where sub-optimally initialised simulations achieve their science goals, reducing runtimes by improving initialisation methods has many benefits. Shorter simulation runtimes accelerate innovation, use research funding more efficiently, maximise the total amount of computational research that can be accomplished, and reduce the environmental impact of astrophysics research through energy and carbon savings.

A novel approach to initialisation involves previewing the linear dynamics of the system using techniques that are less computationally expensive than a full simulation. Recently, demand for less expensive previews has led to a renewed interest in spectroscopic techniques. In \citet{Goedbloed2018a,Goedbloed2018b}, the natural oscillations and instabilities of an equilibrium are quantified by means of a spectral web technique whilst the \legolas{} code \citep{Claes2020,DeJonghe2022,Claes2023} aims to quantify these eigenmodes by reducing the linearised equations to an eigenproblem, based on the method described by \citet{Nijboer1997}.

With such a tool, it now becomes quite easy to initialise non-linear simulations with a superposition of an equilibrium and one or more of its linear eigenmodes, as done in \citet{DeJonghe2025}, perturbing all macroscopic quantities in a consistent way. In a system where all modes are perturbed with small amplitudes, the linear mode with the largest growth rate is expected to become dominant at later stages within the linear evolution. The early evolution, on the other hand, might be affected by the exact form of the initial perturbation, as might the details of the non-linear phase, which typically involves interaction between different modes or secondary instabilities.

In this work, we show that this is indeed the case, and that the choice of initial condition strongly affects how fast the simulation evolves towards a state governed by the instability, allowing us to reduce the computation time by initialising simulations with an appropriate linear solution. Here, we show this for three configurations, each representing a distinct kind of non-linear evolution. For Case 1 we consider a flow-sheared fluid interface, unstable to the Kelvin-Helmholtz instability, as found, for example, in flanks of planetary magnetospheres \citep{Hasegawa2004,Johnson2014}, and planetary atmospheres and oceans \citep{Vallis2017}. This instability drives the system towards a turbulent state, for which we show that its statistical properties are unaltered by this new initialisation method. In Case 2, we investigate the resistive tearing instability \citep{Furth1963} in a Harris current sheet. After the initial formation of plasmoids, two competing, secondary instabilities appear: secondary tearing between plasmoids \citep[sometimes called the plasmoid instability, see e.g.][]{Loureiro2007, Bhattacharjee2009} and plasmoid coalescence. We show that the final state is unaffected by the choice of initial perturbation, but that this choice does affect the interplay between the secondary instabilities. Finally, Case 3 presents a magnetised plasma interface, such as may be found in molecular clouds \citep{Berne2010} or the solar corona, for instance in coronal mass ejections \citep{Foullon2011} and oscillating coronal loops \citep{Howson2017}. Our choice of parameters in this case ensures that the interface is linearly unstable to the Kelvin-Helmholtz instability, but that the non-linear behaviour counteracts the growing instability, stabilising the system again \citep{Frank1996}.

\section{Methodology}
In this work, we used the \amrvac{} code \citep{Porth2014} to perform two-dimensional ($2$D) simulations of both a hydrodynamic and magnetohydrodynamic flow-sheared density gradient (fluid and plasma interface), as well as a magnetohydrodynamic Harris current sheet. Though all three configurations represent one-dimensionally varying, force-balanced states, they are unstable and thus we perturbed these equilibria to trigger their evolution. In our main approach, we computed the eigenmode spectrum of each equilibrium with the \legolas{} code \citep{Claes2020} and added a (superposition of) linear eigenmode(s) to the equilibrium with a predetermined energy, rather than introducing a (semi-)arbitrary analytic perturbation or random noise in a single (scalar or vector) variable, as is common practice \citep[see e.g.][respectively]{Sen2023, Beg2022}.

For an equilibrium varying only along the $x$-direction, the symmetry in the other two directions allows for the use of Fourier modes in the $y$- and $z$-direction, so \legolas{} assumes that the linear solution of each perturbed quantity $f_1$ takes the form
\begin{equation}\label{eq:perturbation}
    f_1(\bfx, t) = \hat{f}_1(x)\,\exp\left[ \im (\bfk\cdot\bfx - \omega t) \right],
\end{equation}
where $\bfx$ and $t$ represent the position vector and time, respectively, $\omega$ is the angular frequency, and $\bfk = k_y\,\ey+k_z\,\ez$ is the wave vector. We note that Eq. \ref{eq:perturbation} does not include a $k_x$ in the wave vector nor assumes anything about the boundary conditions in the $x$-direction, though the application of periodic boundary conditions in that direction may establish an effective wave number $k_x$ in the solution of $\hat{f}_1(x)$. For any given $\bfk$, \legolas{} determines the pairs of frequency $\omega$ and amplitude vector $\hat{\bm{f}}_1(x)$, in which each component represents a perturbed quantity, using a finite element discretisation combining quadratic and cubic elements \citep{Claes2020}. Though the general form of $\bfk$ allows both parallel and oblique modes, in this paper we set $k_z = 0$ since the solutions were used to initialise 2D simulations. The use of sufficiently high resolution was ensured with a frequency convergence test for each configuration and inspection of the corresponding eigenfunctions, presented in App. \ref{app:lego-conv}.

Since equilibria are not equally unstable to all wavelengths ($\lambda = 2\pi/|\bfk|$), we first determined the wave number $k_\mathrm{max}$ corresponding to the largest growth rate (approximately). Unstable eigenmodes were calculated by first exploiting the extremely efficient, single-mode \textsf{inverse-iteration} solver in \legolas{} to find the most unstable frequency for each $\bfk$. For $k_\mathrm{max}$, the full spectrum was then computed with \legolas{}'s \textsf{QR-cholesky} solver, including the perturbations associated with each frequency. From this spectrum, modes were selected to act as initial perturbations in simulations, using Eq. \ref{eq:perturbation} at $t = 0.0$ to obtain a solution throughout the simulation domain. A linear interpolation was used in \legolas{}'s direction of variation to sample the solution on \amrvac{}'s grid.

Since we are comparing different perturbation methods, the perturbation was scaled such that the total energy perturbation is identical for each simulation of a given equilibrium. Hence, we calculated the mean kinetic and internal energies,
\begin{equation}\label{eq:energies}
    E_\mathrm{kin} = \frac{1}{A} \iint_\mathcal{D} \frac{\rho\bfv^2}{2}\dA, \quad E_\mathrm{int} = \frac{1}{A} \iint_\mathcal{D} \frac{p}{\gamma-1}\dA,
\end{equation}
with $p = \rho T$, as well as the mean magnetic energy,
\begin{equation}\label{eq:Emag}
    E_\mathrm{mag} = \frac{1}{A} \iint_\mathcal{D} \frac{\bfb^2}{2}\dA,
\end{equation}
if present, where $\mathcal{D}$ represents the 2D simulation domain and $A$ its area. The construction of all initial perturbations, and their scaling to achieve the desired perturbation in the total energy as defined above, are described in detail in App. \ref{app:scaling}. Denoting the mean total energy $E = E_\mathrm{kin} + E_\mathrm{int} + E_\mathrm{mag}$, we further define the relative energy difference
\begin{equation}
    \overline{\Delta E}(t) = \frac{E(t) - E(0)}{E(0)}
\end{equation}
to quantify a simulation's energy conservation up to simulation time $t$ (in code units).

All simulations assumed perfectly conducting wall boundary conditions in the $x$-direction, in line with \legolas{}'s boundary conditions, and periodic boundary conditions in the $y$-direction. The simulations were performed with \amrvac{} using a three-step, strong stability preserving Runge-Kutta (\textsf{ssprk3}) time integration. For the hydrodynamic configuration in Sec. \ref{sec:fluid-khi}, a Total Variation Diminishing Lax-Friedrichs (\textsf{TVDLF}) flux scheme and a second-order symmetric flux limiter, \textsf{minmod} \citep[see e.g.][]{LeVeque2002}, were used; whereas the magnetohydrodynamic simulations of Secs. \ref{sec:harris} and \ref{sec:plasma-khi} were performed with a Harten-Lax-van Leer (\textsf{HLL}) flux scheme and \textsf{WENO5} flux limiter \citep{Jiang1996}. Simulation convergence tests are presented in App. \ref{app:amrvac-conv}. All quantities in this work are presented in their dimensionless form, following the conventions of the \legolas{} code. For both Kelvin-Helmholtz configurations, all units were derived from the unit length $L_0 = 10^9\,\mathrm{cm}$, unit temperature $T_0 = 10^6\,\mathrm{K}$, unit magnetic field $B_0 = 10\,\mathrm{G}$, and mean molecular weight $\mu = 1$. The Harris sheet configuration used the same $L_0$ and $T_0$, but $B_0 = 2\,\mathrm{G}$ and $\mu = 1/2$ instead.

\section{Results}
\subsection{Turbulent evolution: Flow-sheared fluid interface}
\label{sec:fluid-khi}
\begin{figure}
\centering
    \includegraphics[width=0.5\textwidth]{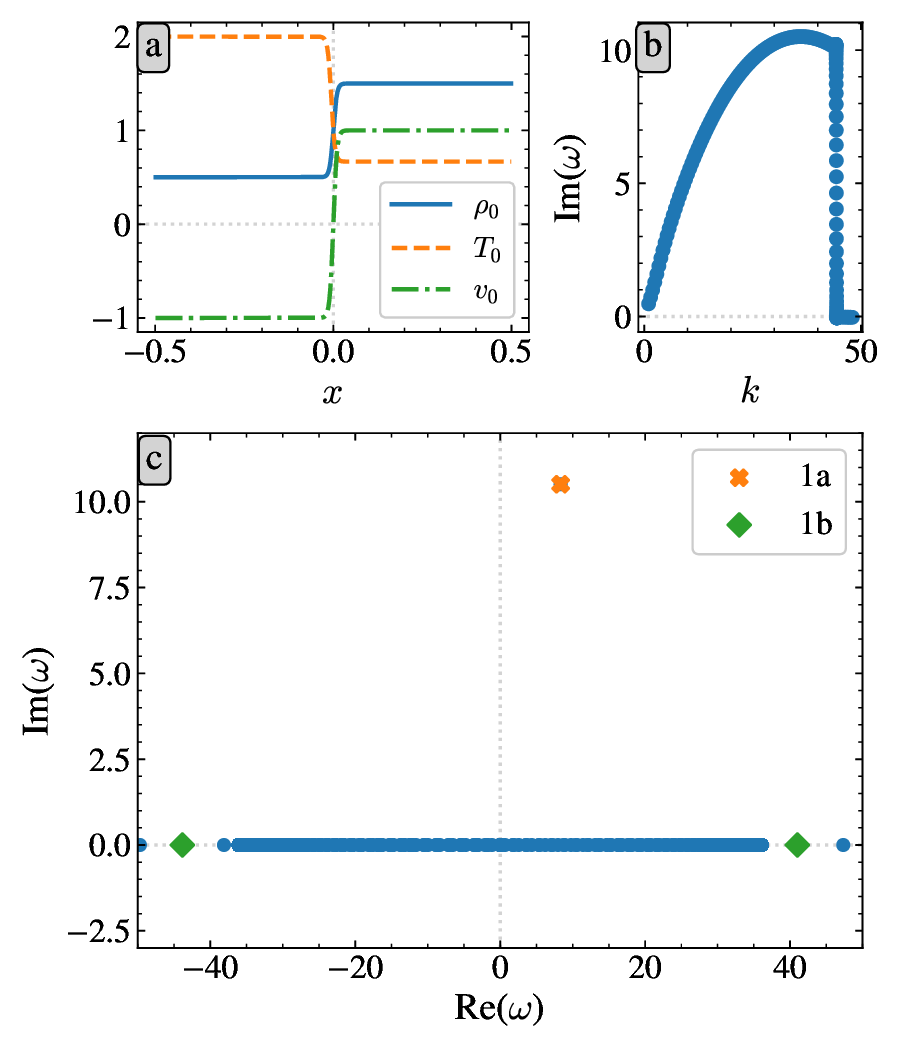}
    \caption{(a) Equilibrium profiles of the flow-sheared fluid interface (Case 1), Eqs. \ref{eq:fluid1}-\ref{eq:fluid2}. (b) KHI growth rate for varying $\bfk = k\,\ey$. (c) \legolas{} spectrum showing the $k = 36.077$ eigenfrequencies of Eqs. \ref{eq:fluid1}-\ref{eq:fluid2} in the complex plane.}
    \label{fig:legolas-khi}
\end{figure}

As a first exploration of the eigenmode perturbation method, we consider a relatively simple setup with only one instability: the Kelvin-Helmholtz instability. This flow-shear driven instability leads to the development of turbulence, for which we then investigate if the choice of initial perturbation affects the turbulent properties. In particular, we consider an ideal, hydrodynamic equilibrium featuring flow shear on a density gradient (Case 1):
\begin{align}
    \rho_0(x) &= 1 + \frac{1}{2} \tanh\left( \frac{x}{a} \right), \label{eq:fluid1} \\
    T_0(x) &= \frac{1}{\rho_0(x)}, \\
    \bfv_0(x) &= v_0 \tanh\left( \frac{x}{a} \right)\ \ey, \label{eq:fluid2}
\end{align}
where $\rho$, $T$, and $\bfv$ represent the density, temperature, and velocity, respectively. The flow speed was set to $v_0 = 1$ and the transition width parameter was set to $a = 0.01$ with $x\in [-0.500, 0.500]$. These equilibrium profiles are shown in Fig. \ref{fig:legolas-khi}a. The use of the ideal gas law $p = \rho T$ implies that the pressure is initially constant.

Due to the presence of flow shear, this equilibrium is unstable to the Kelvin-Helmholtz instability (KHI). Exploiting the low computational cost of \legolas{}, we performed a parameter sweep of the wave number $k$ for wave vector $\bfk = k\,\ey$, employing a centrally accumulated grid as described in \citet{DeJonghe2024} with $p_1 = 0.750$, $p_2 = 0$, $p_3 = 0.001$, and $p_4 = 2.500$, resulting in $229$ grid points on the interval $x\in [-0.500, 0.500]$. The growth rate of the most unstable mode is shown in Fig. \ref{fig:legolas-khi}b for each $k$. The maximum occurs around $k_\mathrm{max} \simeq 36$, for which the central part of the eigenfrequency spectrum is shown in Fig. \ref{fig:legolas-khi}c, with the oscillatory part on the horizontal axis and the growth rate on the vertical axis.

To be close to a multiple of the most unstable wavelength, the domain width in the $y$-direction was set to $L_y = 6\lambda = 12\pi/k = 1.045$ for $k = 36.077$, with $y \in [-0.522, 0.522]$. The fastest growth is achieved by initialising a simulation with the KHI eigenmode with this wave number. Additionally, as pointed out by \citet{Baty2003}, KH unstable configurations evolve to turbulent states once the symmetry between adjacent KH vortices is broken. To avoid reliance on numerical errors to break this symmetry, and thus further hasten the onset of turbulence, we included modes with various wavelengths to ensure asymmetry, however small, between the developing vortices. Hence, modes of wavelength $\lambda$ satisfying $L_y = n\lambda$ with $n = 1, 2, 3, 4, 5, 6$ were included in the initial perturbation such that both the fastest growing wavelength ($n=6$) and longer wavelengths are present to break vortex symmetry. The included wave numbers are thus $k_1 = 6.013$, $k_2 = 12.026$, $k_3 = 18.039$, $k_4 = 24.051$, $k_5 = 30.064$, and $k_6 = 36.077$. For the first simulation (Case 1a), the (dominant) KHI mode (orange cross in Fig. \ref{fig:legolas-khi}c, showing the spectrum of $k_6$) was selected for each wave number for the initial perturbation, whereas Case 1b started from a superposition of a forward and a backward propagating wave for each wave number (green diamonds). The frequencies for each wave in Case 1b can be found per wave number in Table \ref{tab:waves}. Case 1c acts as a reference case and was initialised with a random noise in the velocity. All details of the perturbation construction can be found in App. \ref{app:scaling} alongside an overview of all simulations and their initialisation methods in Table \ref{tab:overview}.

In all three cases, the amplitude of the total perturbation was set such that the mean total energy perturbation is $10^{-9}\,E_0$. For Cases 1a and 1b, modes with the same wave number were added together with equal maximal perturbations in $v_y$ and first scaled to an energy perturbation of $10^{-10}\,E_0$, as described in App. \ref{app:scaling}, before superposing them all, including a phase shift factor $\exp(\im (6-n)\pi/36)$ for wave number $k_n$ ($n=1,\dots,6$), and rescaling the total perturbation to the desired energy perturbation. For Case 1c, $\bfv_1$ in Eq. \ref{eq:noise} was rescaled to satisfy $0 \leq |\bfv_1| \lesssim 7.727\times 10^{-5}$ to achieve the desired expected energy perturbation. The simulations were run on a domain $x \in [-0.500,0.500]$ by $y \in [-0.522,0.522]$ discretised uniformly in each direction in $4096$ and $4224$ cells respectively.

\begin{table}[]
    \centering
    \caption{Wave numbers and corresponding frequencies included in the superposition of perturbations in Case 1b.}
    \label{tab:waves}
    \begin{tabular}{c|cc}
        $k$ & $\omega_1$ & $\omega_2$ \\
        \hline
        $6.013$ & $-11.634$ & $11.737$ \\
        $12.026$ & $-14.626$ & $15.712$ \\
        $18.039$ & $-23.395$ & $25.309$ \\
        $24.051$ & $-26.371$ & $28.316$ \\
        $30.064$ & $-35.112$ & $37.769$ \\
        $36.077$ & $-43.863$ & $40.997$
    \end{tabular}
\end{table}

\begin{figure}
\centering
    \includegraphics[width=0.5\textwidth]{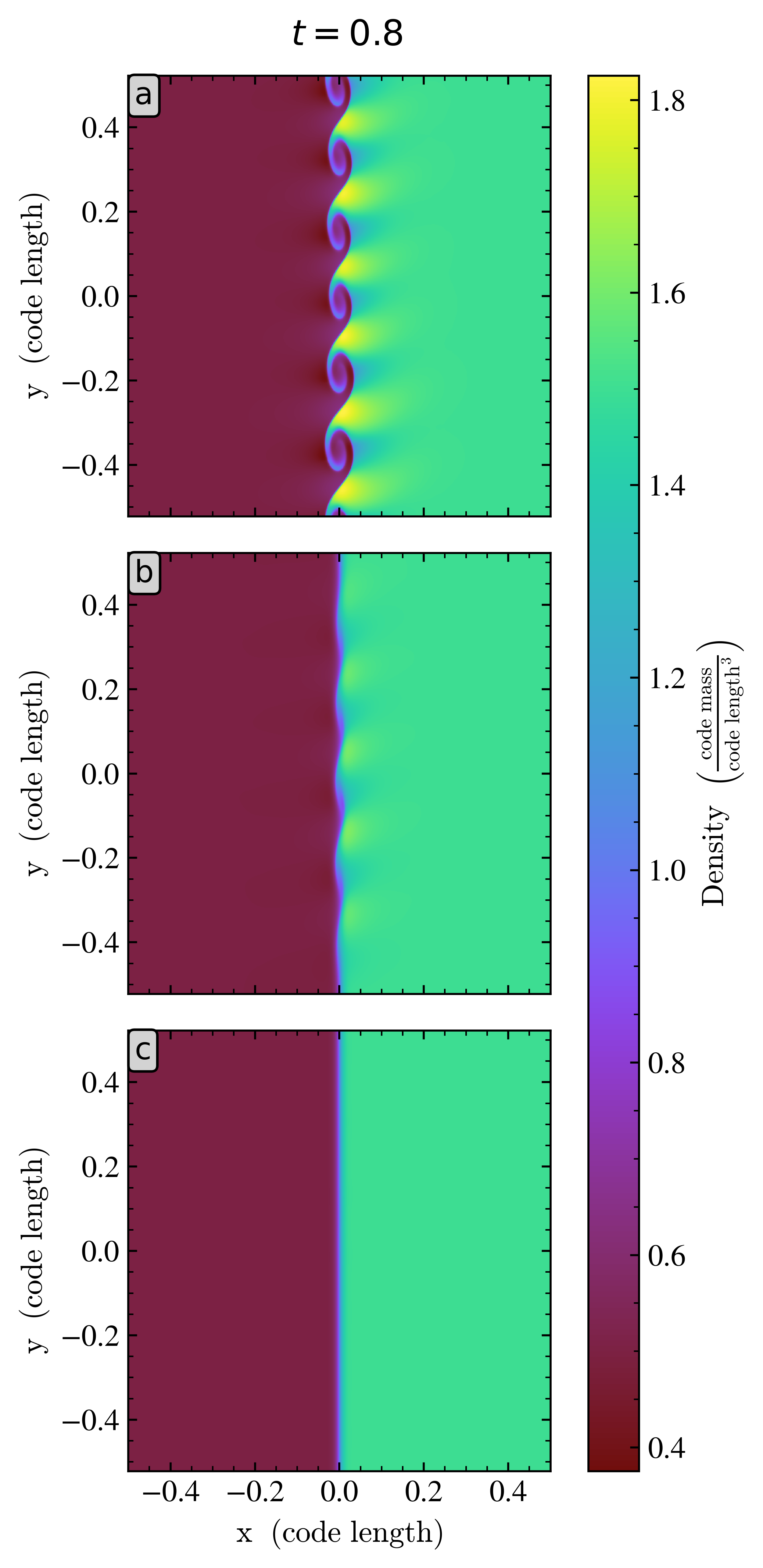}
    \caption{Snapshots of the density at $t = 0.8$ in Case 1 simulations starting from a perturbation with (a) KHI modes; (b) propagating waves; and (c) velocity noise.}
    \label{fig:amrvac-khi}
\end{figure}

\begin{figure}
\centering
    \includegraphics[width=0.5\textwidth]{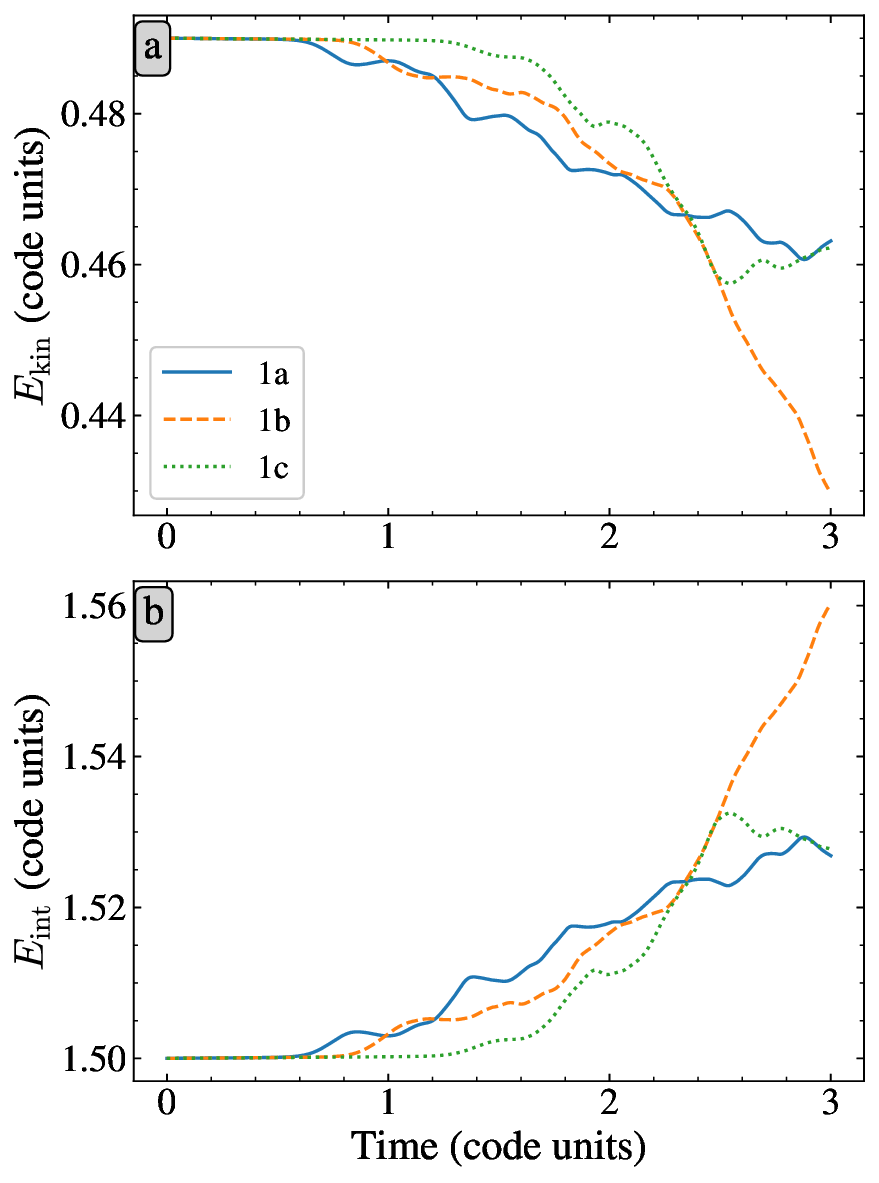}
    \caption{Evolution of the mean (a) kinetic and (b) internal energy in Case 1 simulations starting from a perturbation with (1a) KHI modes, (1b) propagating waves, and (1c) velocity noise.}
    \label{fig:energy-khi}
\end{figure}

For all three simulations, the density is shown at $t = 0.8$ (code time) in Fig. \ref{fig:amrvac-khi}. In Case 1a (Fig. \ref{fig:amrvac-khi}a), the KHI's characteristic swirls at the fluid interface are well-developed. In Case 1b (Fig. \ref{fig:amrvac-khi}b), on the other hand, the instability is only starting to develop at this time whereas Case 1c (Fig. \ref{fig:amrvac-khi}c) is not showing any clear signs of the instability yet. This implies that the simulation initialised with the instability reaches later stages of evolution sooner.

\begin{figure}
\centering
    \includegraphics[width=0.5\textwidth]{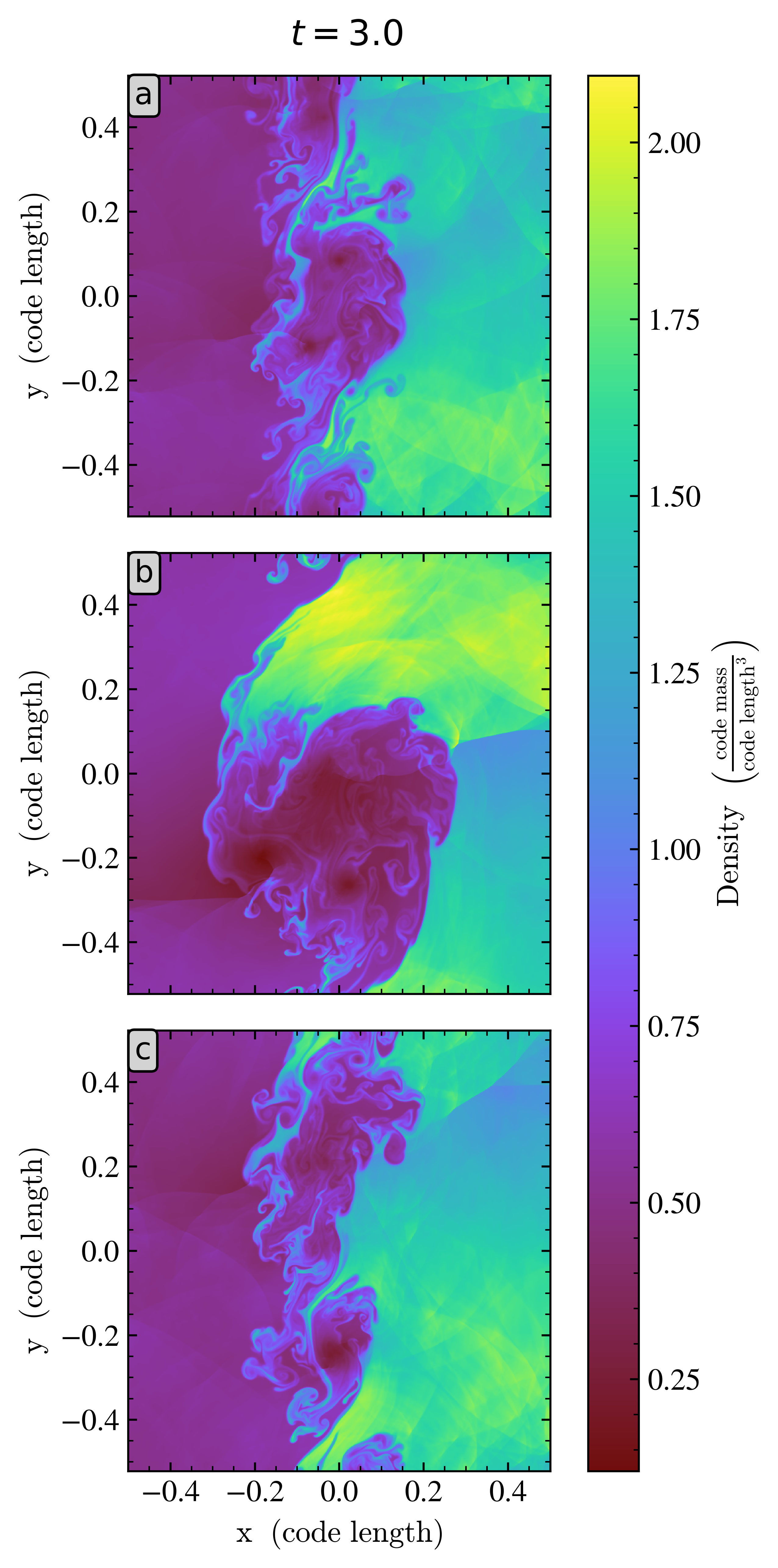}
    \caption{Snapshots of the density at $t = 3.0$ in Case 1 simulations starting from a perturbation with (a) KHI modes; (b) propagating waves; and (c) velocity noise. The associated movie is available online.}
    \label{fig:amrvac-khi2}
\end{figure}

To show that all three simulations behave similarly, and to quantify how fast each simulation evolves, the mean kinetic and internal energies (as defined in Eqs. \ref{eq:energies}) are shown in Fig. \ref{fig:energy-khi} as a function of time. Since the simulation does not include any energy gains or losses, a decrease in kinetic energy is in theory balanced by an equal increase in internal energy. Indeed, quantifying the relative energy differences for $t=3$ (code units) yields $\overline{\Delta E}_{1a}(t=3) = 4.955\times 10^{-12}$, $\overline{\Delta E}_{1b}(t=3) = 4.989\times 10^{-12}$, and $\overline{\Delta E}_{1c}(t=3) = 4.838\times 10^{-12}$. Hence, it suffices to focus our attention on Fig. \ref{fig:energy-khi}a. All three simulations start with a comparable mean kinetic energy, with a small deviation due to the different initial perturbations (only the total energy perturbation is equal between simulations). The kinetic energy decreases in all simulations, with similar initial evolution, though the long-term behaviour of Case 1b deviates from the other two cases due to a difference in vortex merging. We note that this difference may grow from the fact that the three simulations develop differing numbers of vortices initially (6, 5, and 7, respectively). This discrepancy may be due to relaxation of the equilibrium before the perturbations have grown sufficiently to take over the evolution, thus altering the most unstable wave number. Fig. \ref{fig:amrvac-khi2} shows each simulation at its final time, where it is clear that Case 1b evolves more towards one global vortex whereas the other two simulations show remnants of two vortices in the final time (see also the supplementary animation).

Taking $E_\mathrm{kin} = 0.489$ as a threshold value for the start of the non-linear evolution, we find that the early stages of Cases 1b and 1c take $1.294$ and $1.985$ times longer than Case 1a, respectively, to reach this point. Furthermore, for our chosen end time Case 1a spends $22.7\%$ of the simulation time in this early stage, whereas Case 1c remains in this preliminary stage for $45.0\%$ of the simulation's duration. As evidenced by the energy curves and Fig. \ref{fig:amrvac-khi2}, all three simulations eventually evolve towards a state where the fluid layers are mixing thoroughly, destroying the fluid interface, as expected.

\begin{figure}
    \centering
    \includegraphics[width=0.5\textwidth]{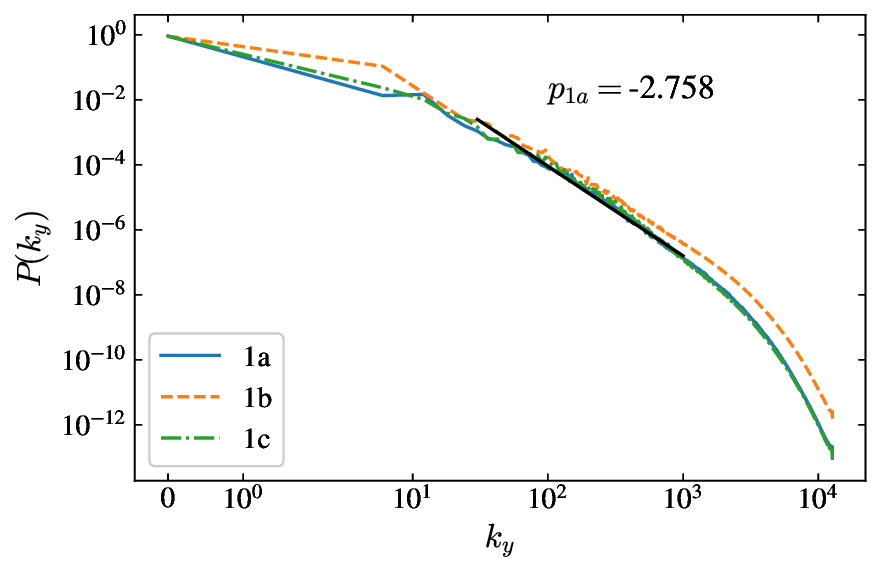}
    \caption{$x$-averaged 1D $k_y$ power spectrum for Case 1 simulations starting from a perturbation with (a) KHI modes; (b) propagating waves; (c) velocity noise; and a power law fit to Case 1a.}
    \label{fig:power-khi}
\end{figure}

As a statistical comparison of the three simulations, we calculated the 1D $k_y$ power spectrum using the velocity magnitude $v$ for all $x$-values in the domain at $t = 3.0$, and took the average over all $x$. For each case, the resulting power spectrum is shown in Fig. \ref{fig:power-khi}. In line with our conclusion from the visual inspection of Fig. \ref{fig:amrvac-khi2}, all three simulations agree well, as desired, obtaining values $p_{1a} = -2.758$, $p_{1b} = -2.613$, and $p_{1c} = -2.855$ for a power law $P(k_y) \sim k_y^p$ fit in the inertial range. This is close to the theoretical value of $-3$ predicted by \citet{Kraichnan1967} and \citet{Batchelor1969} for a 2D enstrophy cascade (some deviation from their result is expected as our initial equilibrium is not homogeneous). Only Case 1b (initialised with propagating waves) shows a limited increase in the power at short and long wavelengths compared to the other two simulations. Hence, the simulations' general behaviour is identical, with eigenmode-initialised simulations spending less time in the preliminary stage of the evolution.

\subsection{Secondary instabilities: Harris current sheet}\label{sec:harris}
\begin{figure}
\centering
    \includegraphics[width=0.5\textwidth]{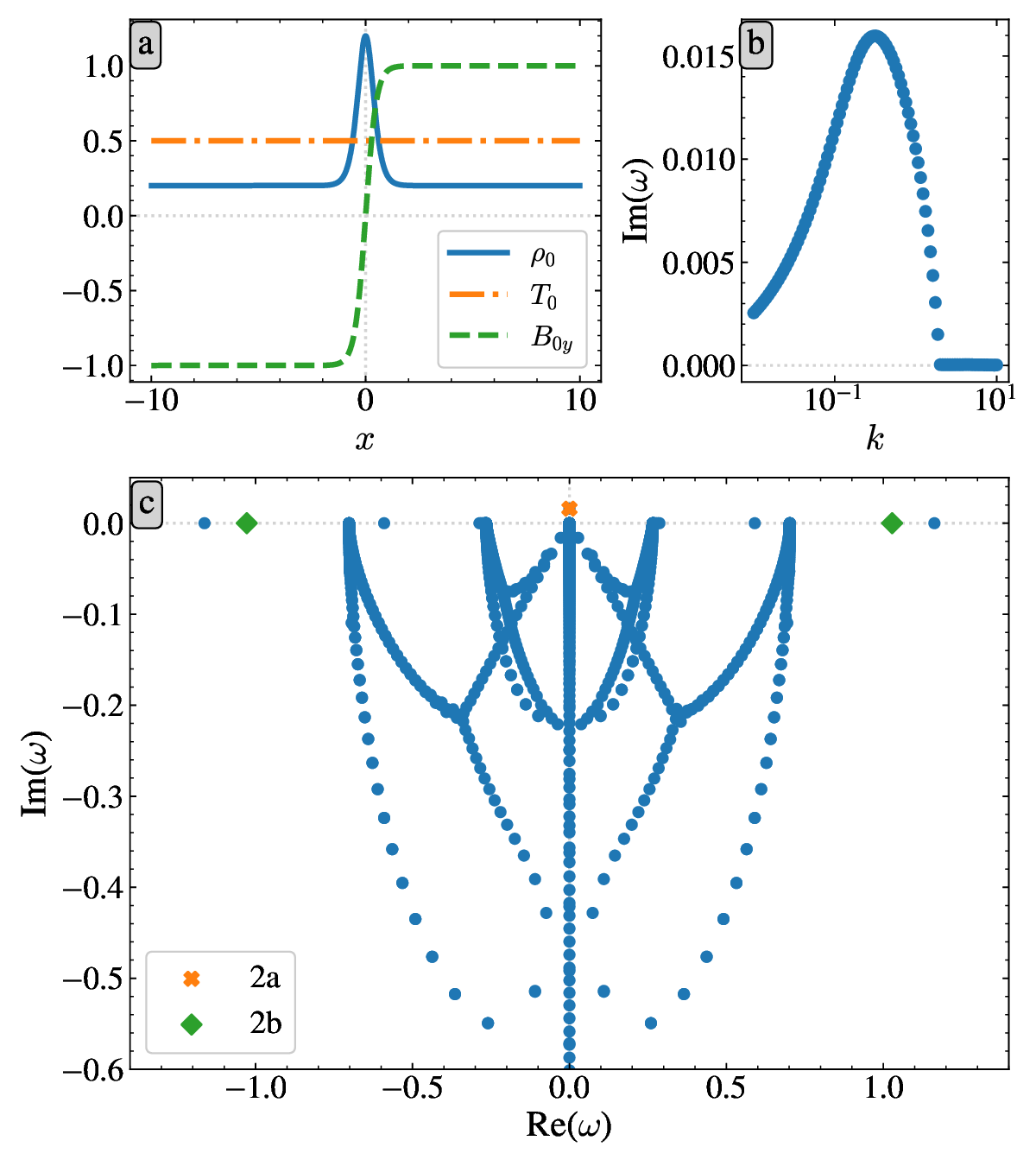}
    \caption{(a) Equilibrium profiles of the static Harris sheet (Case 2), Eqs. \ref{eq:plasma1}-\ref{eq:plasma2}. (b) Tearing growth rate for varying $\bfk = k\,\ey$. (c) \legolas{} spectrum showing the $k = 0.314$ eigenfrequencies of Eqs. \ref{eq:plasma1}-\ref{eq:plasma2} in the complex plane.}
    \label{fig:legolas-harris}
\end{figure}

For this second case we consider a configuration where the instability leads to a new state which is unstable to two types of secondary instabilities. Here, we opt for a current sheet. This structure is unstable to the resistive tearing instability \citep{Furth1963}, causing the current sheet to fragment into magnetic islands, or plasmoids, with smaller-scale current sheets between them. This new configuration is unstable to plasmoid coalescence, but the newly formed current sheets between the plasmoids are simultaneously unstable to secondary tearing \citep[sometimes referred to as the plasmoid instability, see e.g.][]{Loureiro2007,Bhattacharjee2009}. Hence, this case allows us to look at how the initial perturbation affects an evolution with competing secondary instabilities. Specifically, we consider a static Harris current sheet (Case 2) as a magnetohydrodynamic equilibrium,\footnote{This is only an exact equilibrium in ideal MHD. In resistive MHD, it diffuses over a long timescale.}
\begin{align}
    \rho_0 &= \frac{1}{5} + \text{sech}^2 \left(\frac{x}{a}\right), \label{eq:plasma1} \\
    T_0 &= \frac{1}{2}, \\
    \bfv_0 &= \bm{0}, \\
    \bfb_0 &= \tanh\left(\frac{x}{a}\right)\ \ey, \label{eq:plasma2}
\end{align}
where we introduce the magnetic field $\bfb$. Once again, we considered a single parameter choice here, namely $a = 0.500$. Due to the steep $B_{y0}$-gradient in the centre, we used an accumulated grid in \legolas{} as described in \citet{DeJonghe2024}, with $p_1 = 0.200$, $p_2 = 0$, $p_3 = 0.001$, and $p_4 = 5$, resulting in $757$ grid points on the interval $x\in [-10, 10]$. The equilibrium profiles are shown in Fig. \ref{fig:legolas-harris}a.

\begin{figure}
\centering
    \includegraphics[width=0.5\textwidth]{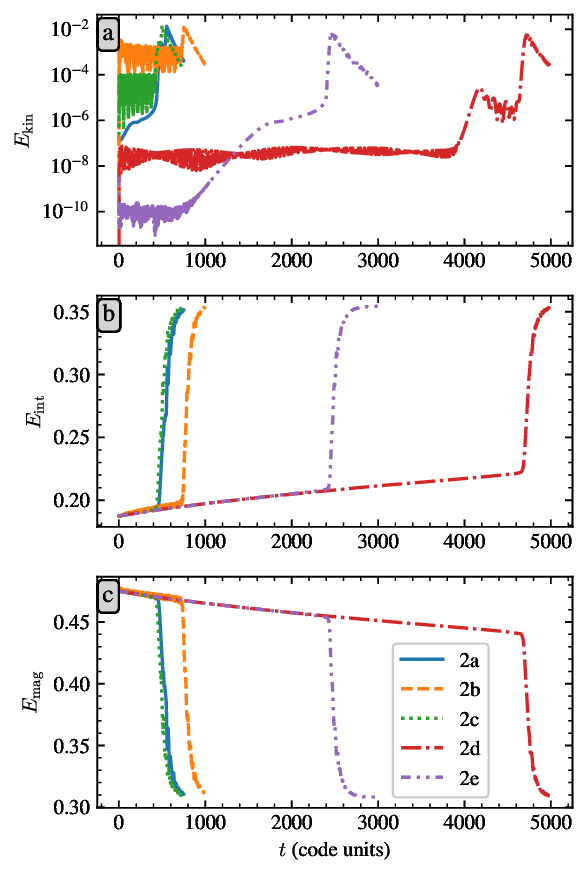}
    \caption{Evolution of the mean (a) kinetic, (b) internal, and (c) magnetic energy in Case 2 simulations starting from a perturbation with (2a) the tearing mode, (2b) a fast wave pair, (2c) a superposition of the tearing mode and a fast wave pair, (2d) the analytic magnetic field perturbation Eqs. \ref{eq:sen-bx}-\ref{eq:sen-by}, and (2e) a velocity noise perturbation.}
    \label{fig:energy-harris}
\end{figure}

\begin{figure*}
\centering
    \includegraphics[width=\textwidth]{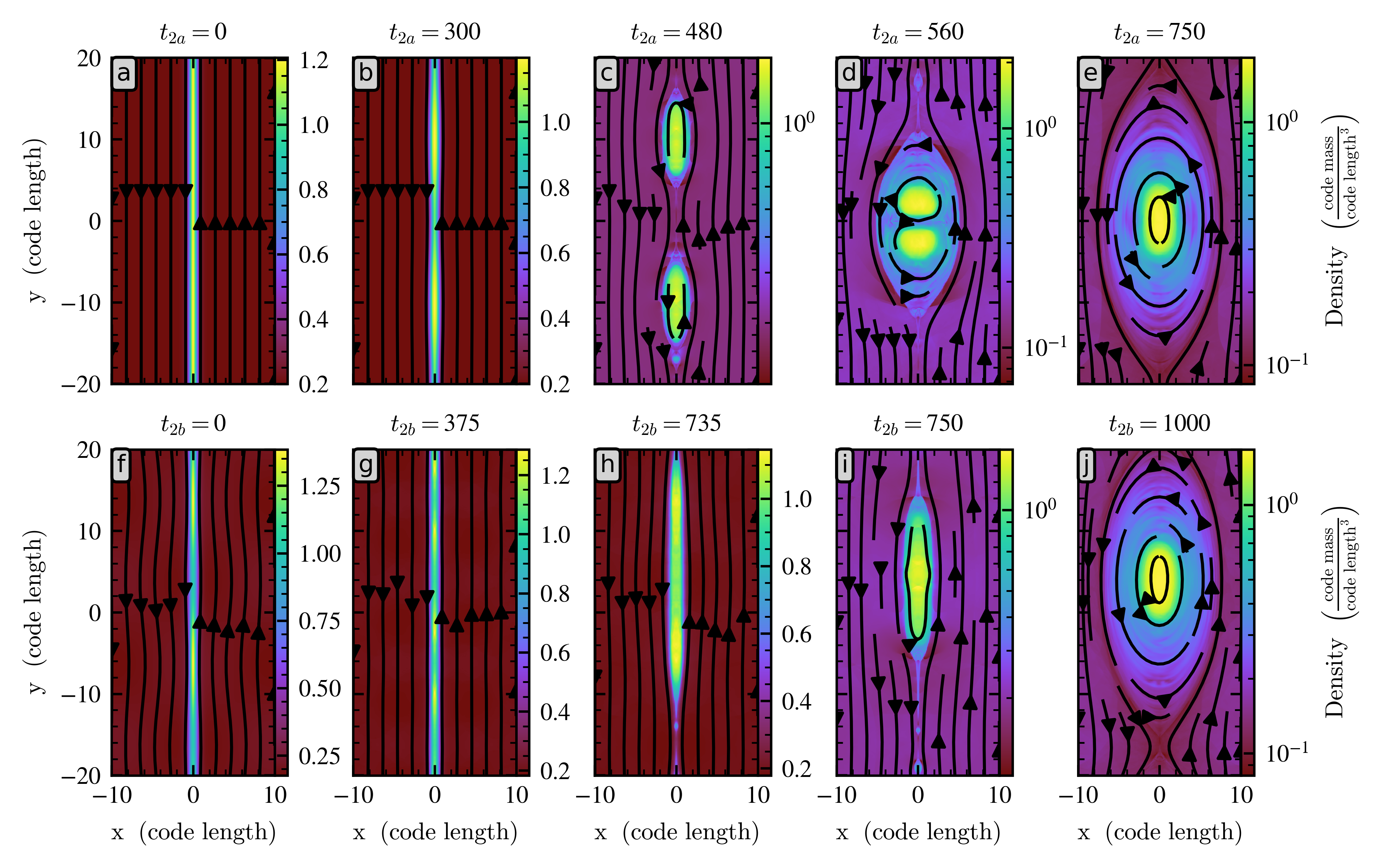}
    \caption{Snapshot of the density and magnetic field lines at various times in (a-e) Case 2a and (f-j) Case 2b, initialised with the fastest tearing mode and a fast wave pair, respectively. The associated movie is available online.}
    \label{fig:amrvac-harris}
\end{figure*}

Due to the magnetic shear, this equilibrium is unstable to the resistive tearing instability \citep{Furth1963} with the inclusion of a non-zero resistivity $\eta = 10^{-4}$. The most unstable wave vector is determined with a \legolas{} parameter sweep of $\bfk = k\,\ey$ as shown in Fig. \ref{fig:legolas-harris}b. The maximum occurs around $k\sim 0.314$, for which the central part of the spectrum is visualised in Fig. \ref{fig:legolas-harris}c.

To study the dynamics of the system with simulations, we defined five cases this time, of which the modes are indicated in the spectrum for the first two cases. For all simulations the bounds in the $y$-direction were set to $y\in[-20.010, 20.010]$ (because $\sim 20.010$ is the wavelength of the fastest growing mode) with $1024$ cells in the $x$-direction and $2048$ in the $y$-direction. The initial perturbation in mean total energy for each case was set to $10^{-8}\,E_0$. For Case 2a, the initial perturbation is the tearing instability itself for $k = 0.314$ (orange cross in Fig. \ref{fig:legolas-harris}c). Case 2b uses a superposition of a forward-backward propagating fast wave pair ($\omega \simeq \pm 1.028-1.914\times 10^{-6}\,\im$, green diamonds) for the same wave number, and Case 2c employs a superposition of the tearing instability (Case 2a) with the fast wave pair from Case 2b. For Cases 2b and 2c, the modes' relative amplitudes were selected in such a way that each mode has an identical maximal $B_y$-perturbation. Finally, we have two reference cases. For Case 2d, the analytic magnetic field perturbation from \citet{Sen2022},
\begin{align}
    \delta B_x = &\frac{L_x}{L_y} A \sin\left( \frac{2\pi x}{L_x} \right) \cos\left( \frac{2\pi y}{L_y} \right), \label{eq:sen-bx} \\
    \delta B_y = &-A \cos\left( \frac{2\pi x}{L_x} \right) \sin\left( \frac{2\pi y}{L_y} \right), \label{eq:sen-by}
\end{align}
was taken with $L_x = 20$ and $L_y = 40.020$ the domain dimensions, and $A \sim 1.029 \times 10^{-4}$ the amplitude ensuring the desired energy perturbation of $10^{-8}\,E_0$. Finally, Case 2e was initiated with a random velocity $\bfv_1$ (Eq. \ref{eq:noise}) in each cell, rescaled to $0 \leq |\bfv_1| \lesssim 2.820 \times 10^{-4}$ to achieve an expected energy perturbation of $10^{-8}\,E_0$. A detailed description of perturbation construction and scaling is given in App. \ref{app:scaling}, with a summary of the five cases in Table \ref{tab:overview}.

\begin{figure}
\centering
    \includegraphics[width=0.5\textwidth]{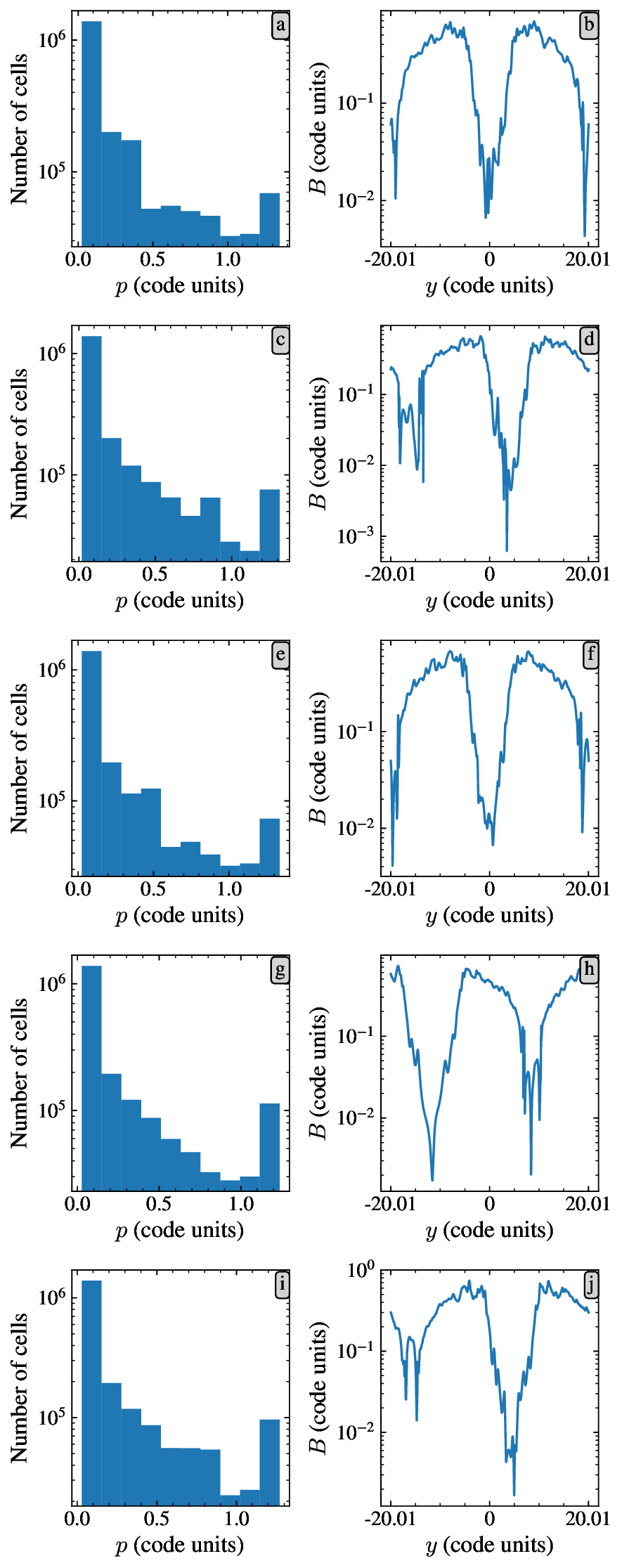}
    \caption{Thermal pressure histograms (left) and magnetic field strength along the current sheet at $x = 0$ from $y = -20.01$ to $y = 20.01$ (right) for (a-b) Case 2a at $t = 750$, (c-d) Case 2b at $t = 970$, (e-f) Case 2c at $t = 690$, (g-h) Case 2d at $t = 4933$, and (i-j) Case 2e at $t = 2708$.}
    \label{fig:stats-harris}
\end{figure}

As in the previous case, we plot the mean kinetic and internal energies in Fig. \ref{fig:energy-harris}, adding the magnetic energy for this MHD case. The simulations conserve the energy to relative energy differences of $\overline{\Delta E}_{2a}(t=750) = -1.122\times 10^{-4}$, $\overline{\Delta E}_{2b}(t=1000) = -1.228\times 10^{-4}$, $\overline{\Delta E}_{2c}(t=750) = -1.143\times 10^{-4}$, $\overline{\Delta E}_{2d}(t=5000) = -7.281\times 10^{-5}$, and $\overline{\Delta E}_{2e}(t=3000) = -8.704\times 10^{-5}$. Additionally, the density and magnetic field lines are shown at various times for Cases 2a and 2b in Fig. \ref{fig:amrvac-harris} (an animation of all evolutions is available as supplementary material). The rightmost panels, Figs. \ref{fig:amrvac-harris}e,j, show the final state that all simulations progress towards. This final state is a single plasmoid (keeping in mind that the boundaries are periodic in the $y$-direction), with a corresponding $E_\mathrm{mag} \sim 0.310$ (code units) as seen in Fig. \ref{fig:energy-harris}c. All simulations eventually end in a single-plasmoid state, as confirmed by Fig. \ref{fig:energy-harris} (and the supplementary animation). To illustrate that the states the simulations progress towards are equivalent, Fig. \ref{fig:stats-harris} shows the thermal pressure histogram and magnetic field strength along $x=0$ from bottom to top for each case at the time the simulation drops below $E_\mathrm{kin} = 4.103\times 10^{-4}$, which is the kinetic energy at the end of simulation 2a. The pressure histograms show a comparable distribution for all simulations, and the magnetic field strength curves are extremely similar, keeping in mind that the domain is periodic in $y$.

Although the energy curves in Fig. \ref{fig:energy-harris} are quite similar, they are not merely temporal translations of each other due to the conversion of magnetic to internal energy by resistive diffusion. This introduces a slope in the mean internal and magnetic energies before plasmoids start to form, and may contribute to the differences in evolution observed in the simulations. In essence, the simulations feature two different evolutions resulting in a single-plasmoid state. Both evolutions are represented in Fig. \ref{fig:amrvac-harris}, with panels a through e showing Case 2a, and panels f through j Case 2b. In the first type of evolution (Fig. \ref{fig:amrvac-harris}a-e), the sheet forms two distinct plasmoids (Fig. \ref{fig:amrvac-harris}c), in accordance with the most unstable wavelength, after which the plasmoids start to merge (Fig. \ref{fig:amrvac-harris}d) into a single plasmoid (Fig. \ref{fig:amrvac-harris}e). Case 2c follows this formation process. In the other three cases (2b, 2d, and 2e), the formation and merging of plasmoids are not clearly separated processes, but happen simultaneously. Fig. \ref{fig:amrvac-harris}f shows how two locations of enhanced density are seeded, but the bent field lines cause oscillations in the sheet (see Fig. \ref{fig:amrvac-harris}g and the supplementary animation). Eventually one elongated plasmoid with two density enhancements is formed (Fig. \ref{fig:amrvac-harris}h), before contracting to a shorter and wider plasmoid (Figs. \ref{fig:amrvac-harris}i). Regardless of the exact formation process, secondary tearing is observed in the short and thinned sheets between plasmoids, generating smaller plasmoids, which can be seen at the bottom of Figs. \ref{fig:amrvac-harris}c,h,i and at the top of Fig. \ref{fig:amrvac-harris}d. These smaller plasmoids are created and merge with the main plasmoid until it reaches its final size.

Delineating the end of the early stage at $E_\mathrm{mag} = 0.440$ (code units) somewhat arbitrarily, we can calculate how long it takes each simulation to leave the early stage. Case 2a reaches this value of $E_\mathrm{mag}$ at $t = 487$ (code units) whilst it takes Cases 2b through 2e respectively $1.540$, $0.947$, $9.563$, and $5.016$ times as long to get there. Notably, the simulation initialised with the superposition of the tearing instability and fast waves (Case 2c) evolves the fastest. All three eigenmode-initialised simulations are more than $3$ times faster than both reference cases.
The tearing eigenmode initialised simulation (2a) is an order of magnitude faster than the simulation initialised with an analytic perturbation (2d). Finally, looking at Fig. \ref{fig:energy-harris}a, Cases 2b, 2c, and 2d show oscillatory behaviour in their kinetic energy, in line with their initialisation with waves. This oscillatory behaviour is damped heavily when the plasmoid formation takes over. On the other hand, Case 2a was not initialised with an oscillating perturbation, and evolves smoothly towards the lowest $E_\mathrm{mag}$ state.

\subsection{Non-linear stabilisation: Flow-sheared plasma interface}\label{sec:plasma-khi}
In this third and final case, we test how the eigenmode initialisation method performs for a configuration that is linearly unstable, but where the non-linear evolution counteracts the growth and eventually stabilises the system. To do so, we again consider a flow-sheared interface described by Eqs. \ref{eq:fluid1}-\ref{eq:fluid2} with $a = 0.01$ in the interval $x\in [-1, 1]$, setting $v_0 = 1.250$, and adding a uniform magnetic field $\bfb_0 = \ey$ ($B_0 = 1$) and plasma resistivity $\eta = 10^{-5}$ (Case 3).

For the textbook flow-sheared plasma interface with a uniform magnetic field but a discontinuous transition between regions with densities and parallel velocities $\rho_1$, $v_1$ and $\rho_2$, $v_2$, respectively, the criterion for linear instability is that the relative flow speed exceeds the root-mean-square (RMS) Alfv\'en speed $v_{\mathrm{A,RMS}}$ \citep[see e.g.][]{Chandrasekhar1961},
\begin{equation}
    (v_1 - v_2)^2 > v_{\mathrm{A,RMS}}^2 = \frac{B^2}{2} \left( \frac{1}{\rho_1} + \frac{1}{\rho_2} \right),
\end{equation}
or expressed differently, that the configuration's `global' Alfv\'en Mach number $M_\mathrm{A} = |v_1-v_2| / v_{\mathrm{A,RMS}}$ exceeds $1$. Neglecting the narrow transition region in our configuration for a moment, we have $\rho_1 = 0.5$, $v_1 = -1.25$ and $\rho_2 = 1.5$, $v_2 = 1.25$ for a relative speed of $|v_1-v_2| = 2.5$ and with $B = 1$ an RMS Alfv\'en speed of $v_{\mathrm{A,RMS}} = 2/\sqrt{3} \simeq 1.155$, or $M_\mathrm{A} \simeq 2.165$. Hence, the configuration is linearly unstable.

Regarding the impact of the magnetic field on the non-linear dynamics, as shown in \citet{Frank1996}, a flow-sheared interface with a large Alfv\'en Mach number (weak magnetic field) evolves in essence hydrodynamically (similar to Case 1 discussed before) until the vortices start to roll over, whereas the magnetic tension exerts a stronger stabilising influence during the non-linear phase as $M_\mathrm{A}$ approaches $1$. If $M_\mathrm{A} \gtrsim 1$, the magnetic field lines that are bent due to the developing KHI, push back and force the disrupted interface to evolve towards a steady state with a nearly laminar flow between both exterior regions, which is similar to the initial configuration, but where the shear layer has broadened until it is stable \citep{Frank1996}.

\begin{figure}
\centering
    \includegraphics[width=0.5\textwidth]{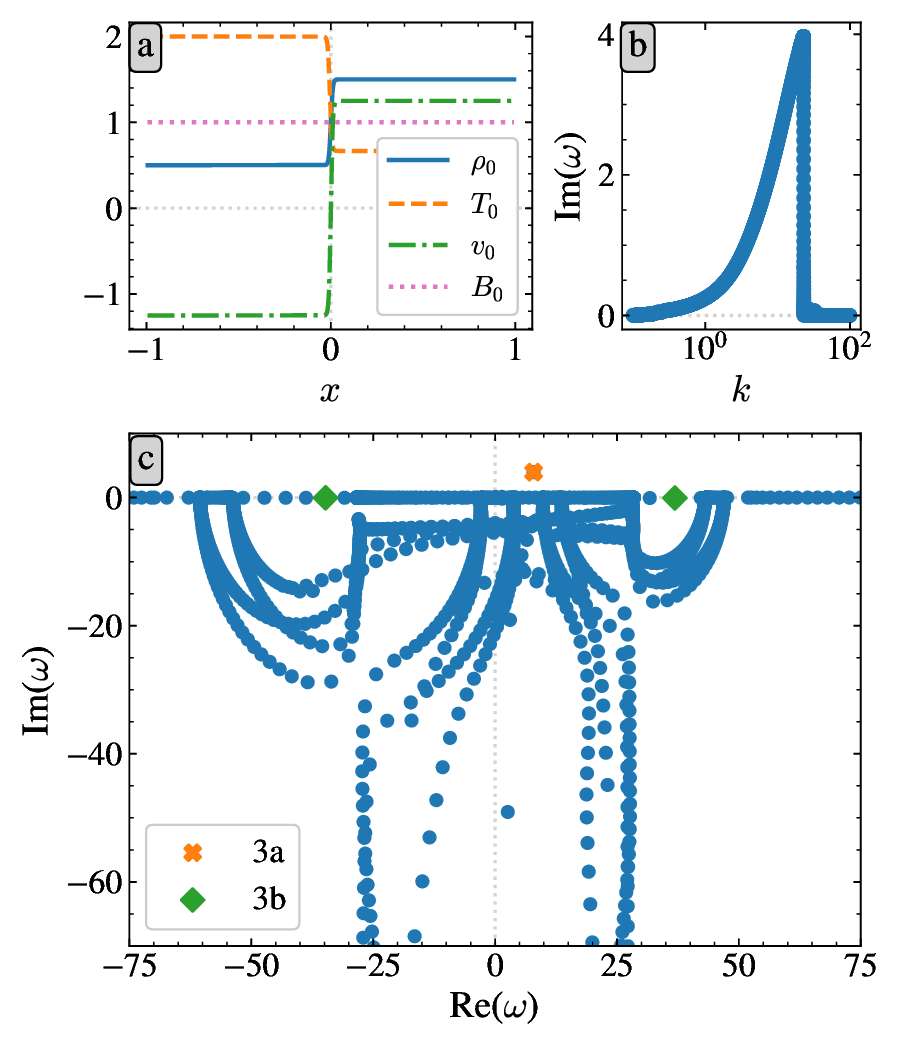}
    \caption{(a) Equilibrium profiles of the flow-sheared plasma interface (Case 3), Eqs. \ref{eq:fluid1}-\ref{eq:fluid2}, with a uniform magnetic field. (b) KHI growth rate for varying $\bfk = k\,\ey$. (c) \legolas{} spectrum showing the $k \simeq 22.730$ eigenfrequencies of Eqs. \ref{eq:fluid1}-\ref{eq:fluid2} with $\bfb_0 = \ey$ in the complex plane.}
    \label{fig:legolas-mhdkhi}
\end{figure}

A parameter sweep with \legolas{} (using an identical grid setup as in the unmagnetised case, for a total of $241$ grid points) reveals that the most unstable wave number is close to $k\sim 22.730$. The equilibrium, parameter sweep, and spectrum for $k \simeq 22.730$ are shown in Fig. \ref{fig:legolas-mhdkhi}. For the subsequent simulation, we set the domain width in the $y$-direction to eight times the corresponding wavelength, $y \in [-1.106, 1.106]$.

Similarly to the fluid KHI (Sec. \ref{sec:fluid-khi}), we defined 5 cases: Case 3a uses magnetised KHI modes, Case 3b uses (weakly damped) propagating waves, Case 3c has both the KHI modes from 3a and the waves from 3b, Case 3d has the shortest wavelength KHI (most unstable mode) from 3a and the longest wavelength waves from 3b, and Case 3e was initialised with velocity noise, as specified in Table \ref{tab:overview} too. Again, a superposition of various wavelengths, as defined in App. \ref{app:scaling}, was used in the initial perturbation to ensure asymmetry between vortices. Since the domain size $L_y$ in the $y$-direction corresponds to eight times the wavelength of the fastest growing mode, the initial perturbations of Cases 3a, 3b, and 3c are superpositions of modes with wavelengths $L_y/n$ for $n=1,\dots,8$, and $n = 1, 8$ for Case 3d. For Case 3a, the dominant KHI mode was selected for each of these wave numbers. For Case 3b, the approximate frequencies of the selected damped modes can be found per wave number in Table \ref{tab:waves-mhd}. In Cases 3a, 3b, 3c, and 3d, modes with the same wave number were added together with equal maximal perturbations in $v_y$ before scaling them to an energy perturbation of $10^{-9}\, E_0$. Subsequently, each wave number's perturbations were phase-shifted with a factor $\exp(\im (8-n)\pi/36)$, after which the perturbations were summed and the total perturbations were scaled to an energy perturbation of $10^{-8}\, E_0$. Case 3e was also initialised with an expected energy perturbation of $10^{-8}\, E_0$.

\begin{table}[]
    \centering
    \caption{Wave numbers and corresponding frequencies included in the superposition of perturbations in Case 3b.}
    \label{tab:waves-mhd}
    \begin{tabular}{l|cc}
        $k$ & $\omega_1$ & $\omega_2$ \\
        \hline
        $2.841$ & $-12.995-1.318\times 10^{-4}\,\im$ & $12.368-8.574\times 10^{-5}\,\im$ \\
        $5.682$ & $-16.613-4.483\times 10^{-4}\,\im$ & $15.362-1.034\times 10^{-4}\,\im$ \\
        $8.524$ & $-28.278-4.833\times 10^{-4}\,\im$ & $21.274-1.626\times 10^{-4}\,\im$ \\
        $11.365$ & $-31.854-2.087\times 10^{-3}\,\im$ & $29.145-3.387\times 10^{-4}\,\im$ \\
        $14.206$ & $-41.296-3.266\times 10^{-3}\,\im$ & $35.148-4.662\times 10^{-4}\,\im$ \\
        $17.047$ & $-53.645-1.773\times 10^{-3}\,\im$ & $41.228-6.039\times 10^{-4}\,\im$ \\
        $19.889$ & $-29.151-2.760\times 10^{-2}\,\im$ & $31.332-6.527\times 10^{-2}\,\im$ \\
        $22.730$ & $-34.773-3.325\times 10^{-2}\,\im$ & $36.874-7.243\times 10^{-2}\,\im$
    \end{tabular}
\end{table}

\begin{figure}
\centering
    \includegraphics[width=0.41\textwidth]{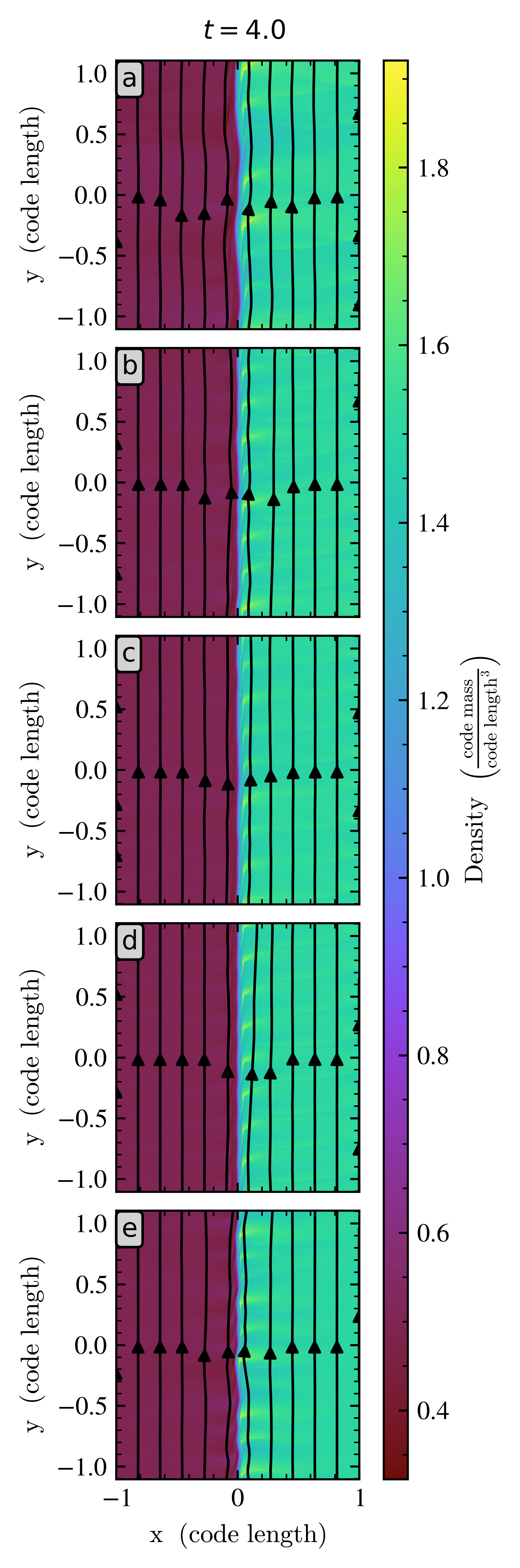}
    \caption{Snapshots of the density, with annotated magnetic fieldlines, at $t = 4.0$ in Case 3 simulations starting from a perturbation with (a) KHI modes; (b) propagating waves; (c) KHI modes and propagating waves; (d) a short wavelength KHI mode and long wavelength waves; and (e) velocity noise.}
    \label{fig:amrvac-mhdkhi}
\end{figure}

In this configuration and for this choice of parameters, the simulations initially develop a displacement of the plasma interface, which first appears in Case 3a, followed in order by 3c, 3d, 3b, and finally 3e. This can be seen in Fig. \ref{fig:amrvac-mhdkhi}, where all cases show a displacement of the interface and smaller structures therein at $t = 4.0$.

\begin{figure}
\centering
    \includegraphics[width=0.5\textwidth]{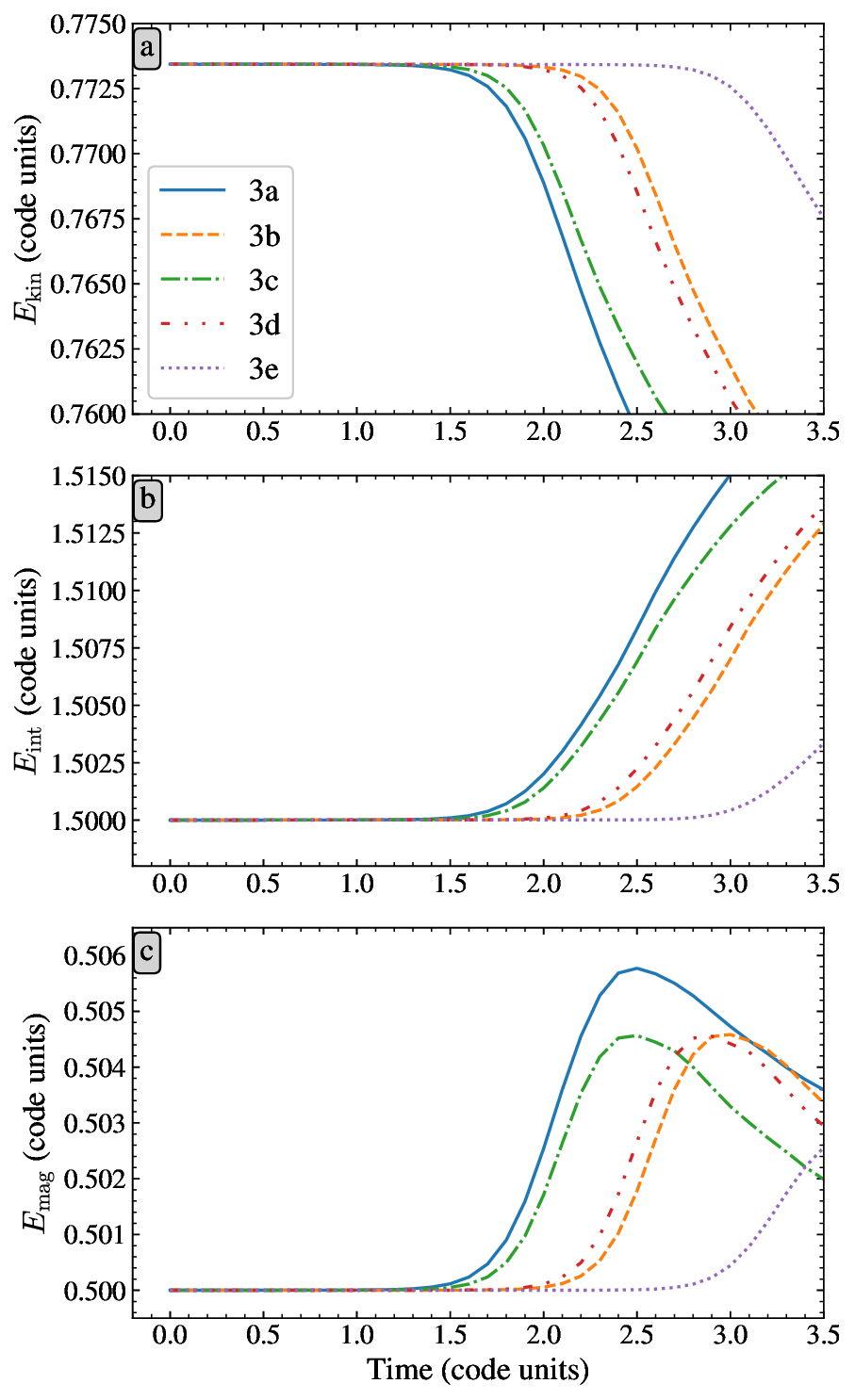}
    \caption{Evolution of the mean (a) kinetic, (b) internal, and (c) magnetic energy in Case 3 simulations starting from a perturbation with (3a) KHI modes, (3b) propagating waves, (3c) KHI modes and propagating waves, (3d) a short wavelength KHI mode and long wavelength waves, and (3e) velocity noise.}
    \label{fig:energy-mhdkhi-zoom}
\end{figure}

The difference in onset time can be seen in the energy curves in Fig. \ref{fig:energy-mhdkhi-zoom}. In the kinetic energy (Fig. \ref{fig:energy-mhdkhi-zoom}a), each curve features a steep decrease as the interface is perturbed, which goes paired with a peak in the magnetic energy (Fig. \ref{fig:energy-mhdkhi-zoom}c) as the magnetic field lines are bent due to their frozen-in state in the displaced plasma. Again imposing a threshold on the kinetic energy, $E_\mathrm{kin} = 0.770$, below which we considered the evolution to have truly commenced, we find that it takes Cases 3b, 3c, 3d, and 3e, respectively, $1.30$, $1.05$, $1.25$, and $1.65$ times as long to exit the early stage compared to Case 3a. As expected the fastest evolution occurs in Case 3a, initialised with the unstable modes, whilst the slowest is Case 3e, initialised with velocity noise. The remaining three cases have onset times between those of 3a and 3e.

\begin{figure}
\centering
    \includegraphics[width=0.5\textwidth]{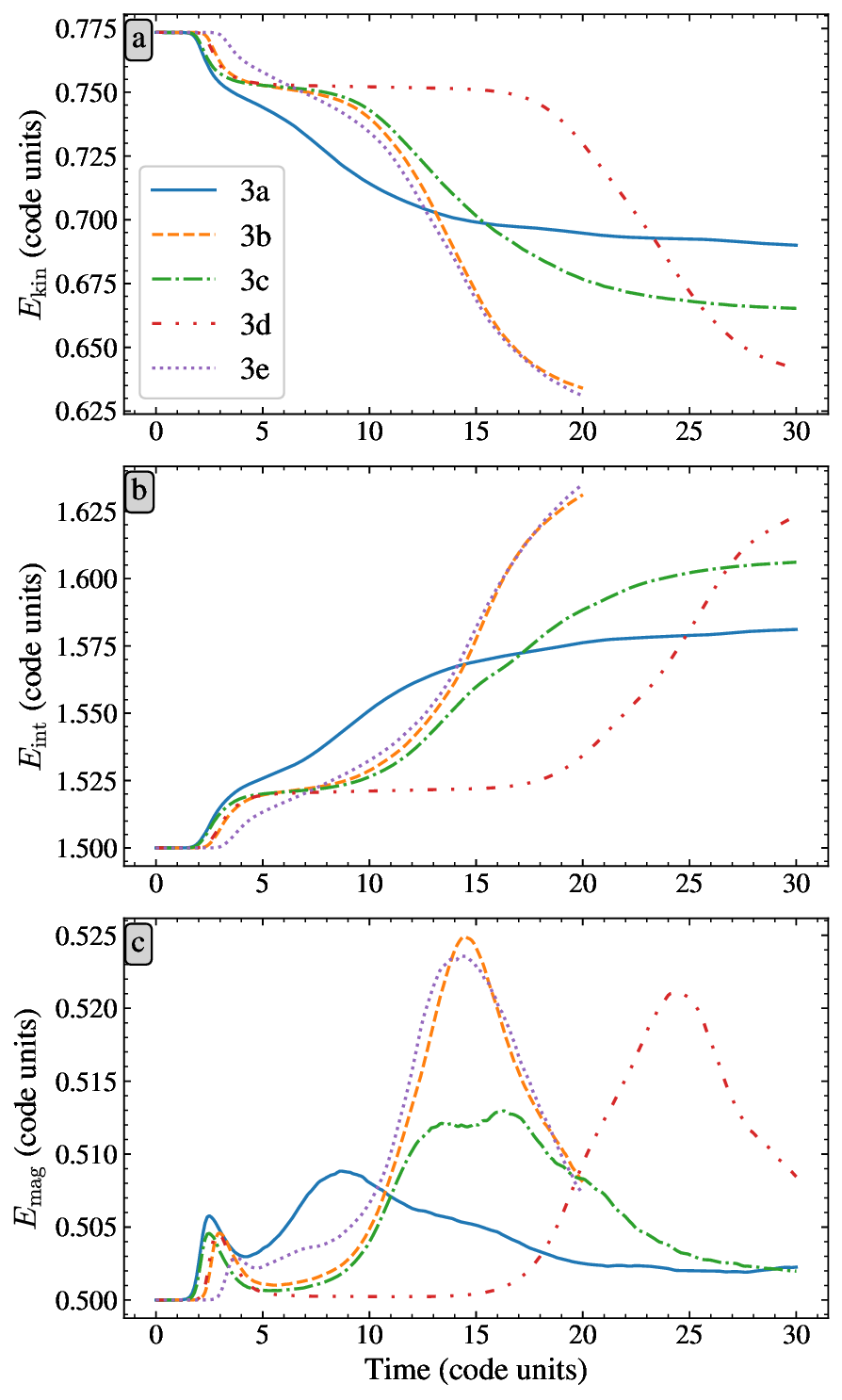}
    \caption{Long-term evolution of the mean (a) kinetic, (b) internal, and (c) magnetic energy in Case 3 simulations starting from a perturbation with (3a) KHI modes, (3b) propagating waves, (3c) KHI modes and propagating waves, (3d) a short wavelength KHI mode and long wavelength waves, and (3e) velocity noise.}
    \label{fig:energy-mhdkhi}
\end{figure}

After the initial stage, the instability is suppressed by magnetic tension, which reduces both the magnetic energy and the rate of decrease in kinetic energy, until the interface perturbation amplitude is smaller again, as seen in Fig. \ref{fig:amrvac-mhdkhi2}, and develops a new steady flow along the interface (see supplementary animation). Inspecting the energy curves past the initial stage (Fig. \ref{fig:energy-mhdkhi}) reveals that the initial perturbation plays an important role in determining the non-linear evolution, with growing discrepancies between simulations. Despite all simulations sharing a similar linear stage, only Case 3b follows the reference Case 3e throughout the entire simulation, forming a single shock in the dense area. Case 3d develops the same structure, but with a significant delay compared to Cases 3b and 3e, with two regimes clearly separated by a plateau in the energies. The biggest outlier is Case 3a, initialised with KHI modes only, and Case 3c ends up between Case 3a and the others in kinetic and internal energy. This implies that whilst the simulations are originally dominated by the short wavelength KHI modes, the long wavelength waves become important in the non-linear stage and govern the long term evolution. Notably, Case 3e lacks the plateau observed in Cases 3b, 3c, and 3d, indicating a smooth transition from KHI to long wavelength wave domination, whereas this does not occur in Case 3a because these waves were not seeded. This behaviour points to the importance of the relative amplitudes of modes in the initial perturbation in non-linearly complex evolutions such as this one, and could be used to study how energy is redistributed across modes and wavelengths in non-linear interactions. All five simulations conserve the energy similarly well with relative energy differences of $\overline{\Delta E}_{3a}(t=20) = -1.778 \times 10^{-5}$, $\overline{\Delta E}_{3b}(t=20) = -7.869 \times 10^{-5}$, $\overline{\Delta E}_{3c}(t=20) = -2.607 \times 10^{-5}$, $\overline{\Delta E}_{3d}(t=20) = -6.560 \times 10^{-5}$, and $\overline{\Delta E}_{3e}(t=20) = -9.299\times 10^{-5}$.

\begin{figure}
\centering
    \includegraphics[width=0.41\textwidth]{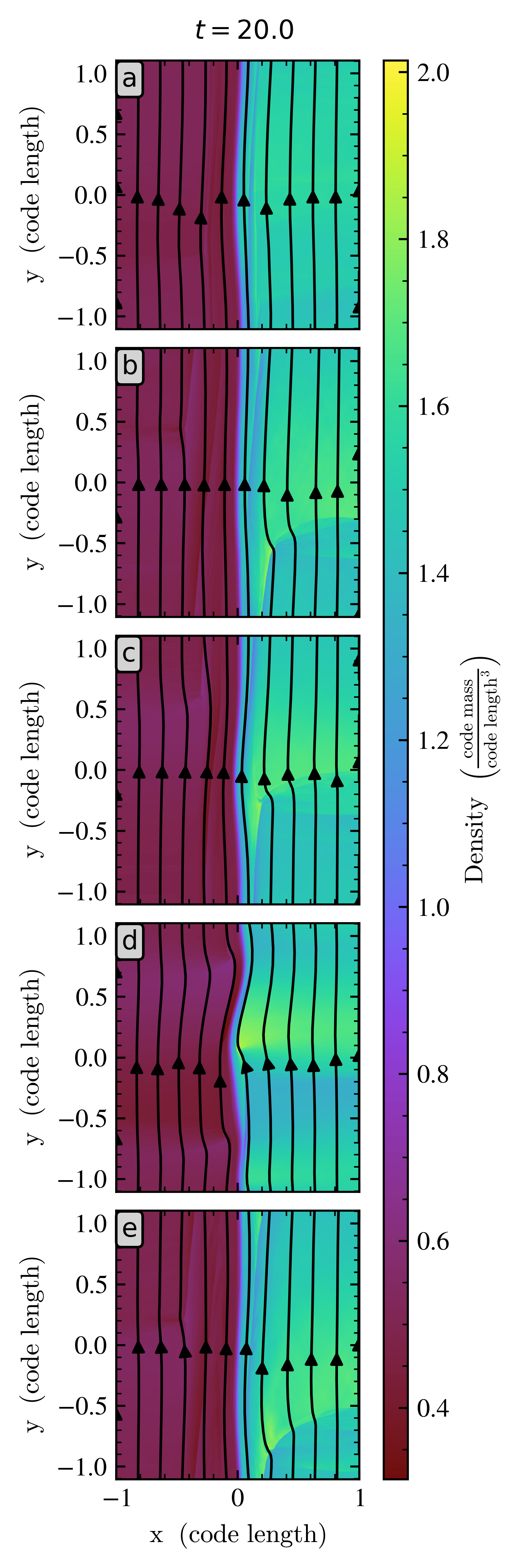}
    \caption{Snapshots of the density, with annotated magnetic fieldlines, at $t = 20.0$ in Case 3 simulations starting from a perturbation with (a) KHI modes; (b) propagating waves; (c) KHI modes and propagating waves; (d) a short wavelength KHI mode and long wavelength waves; and (e) velocity noise. The associated movie is available online.}
    \label{fig:amrvac-mhdkhi2}
\end{figure}

\section{Conclusion}
In hydrodynamic and magnetohydrodynamic simulations, it is common practice to perturb unstable equilibria with analytically prescribed perturbation profiles or random noise to start the evolution. In this work, it was shown that the time spent in the linear growth phase of $2$D simulations can be reduced significantly by initialising with a superposition of linear eigenmodes that includes the fastest growing unstable mode. Comparing Case 2d to Case 2c shows that the time spent in the linear stage can be reduced by up to an order of magnitude, compared to simulations using traditional perturbations, although the factor is problem-dependent as demonstrated by the three cases in this work.

Saving numerical resources by adopting eigenmode initialisation has many general benefits. The saved resources can be allocated elsewhere, for instance to increasing resolution, following the simulation dynamics further in time, or running larger parameter studies. Shorter simulation runtimes also accelerate research, use funding more efficiently, and reduce financial and environmental costs of simulation research.

The new technique is furthermore able to improve the quality of computer experiments. Shortening the linear growth phase makes it more likely that the unstable modes that are detected in the simulation correspond to the unstable modes of the initial state, since there is less time for physical or numerical diffusion to operate. The simulations are also cleaner in the sense that traditional perturbation methods often excite a superposition of waves, which obscure the growth of the fastest growing linear mode until later times. Eigenmode initialisation therefore provides significant benefits for instability studies by providing better guarantees that the desired modes are being studied and by improving data on the growth phase, such as more accurate linear growth rates.

Eigenmode initialisation has the potential to enable entirely new research studies. One major advantage is that instability is achieved earlier not only in wall clock time, but also in simulation time. As noted in the Introduction, numerical experiments sometimes have an intrinsic lifetime determined by dynamics that operate on a long timescale. In such setups, commencing the non-linear dynamics of interest earlier allows the desired phenomenon to be studied for longer, and hence with better statistics. This is applicable, for example, to the authors' own work on self-generated turbulent magnetic reconnection.

On an operational level, each of the three cases presented offers further insight into how to get the most out of this initialisation method. In the flow-sheared fluid interface of Case 1, the inclusion of several wavelengths in the initial perturbation proves critical to obtaining fast symmetry breaking between Kelvin-Helmholtz vortices to develop turbulence. Case 2, which simulates plasmoid formation in a Harris current sheet, reveals that including waves alongside the fastest growing mode can further accelerate the onset, and that the specifics of the later evolution due to secondary instabilities depends on the initial perturbation. Lastly, the non-linearly stabilised plasma interface of Case 3 highlights the importance of including modes in the initial perturbation whose behaviour becomes relevant in the non-linear stage.

The combined experience of the case studies demonstrates that eigenmode initialisation offers a second significant advantage for computer experiments, namely, improved control over non-linear processes. By actively setting the mix of modes included in the initial perturbation, one gains more precise control over symmetry breaking, secondary instability, and slower growing modes that become important in the later dynamics. In one specific scientific application, this advantage offers a new opportunity to rigorously examine the competition between different secondary instabilities, and we plan future work extending eigenmode initialisation to 3D to examine the competition of coalescence and kink instabilities of flux ropes formed by 3D tearing instabilities in reconnecting current layers, which is motivated by the switch-on of fast magnetic reconnection in the solar corona.

In summary, this eigenmode initialisation method provides substantial numerical savings, highlights the dynamics driven by a user-defined selection of modes, and enables clearer separation of short-term phenomena from other dynamics occurring on a long timescale.

\section*{Data availability}
The results in this work were obtained with \legolas{} v2.1.1 and \amrvac{} v3.1, which are available at \url{https://legolas.science} and \url{https://amrvac.org}, respectively. The analysis and visualisation of simulation data was handled with \textsf{yt} \citep[\url{https://yt-project.org}]{Turk2011}.

\begin{acknowledgements}
The authors thank the referees for suggestions that improved the paper. JDJ thanks R. Keppens and D. Maci for fruitful discussions regarding Kelvin-Helmholtz simulations and power spectra of turbulent states. AJBR thanks A. Hillier for conversations about non-linear stabilisation of linear MHD instabilities by magnetic tension. The authors acknowledge funding by the UK's Science and Technology Facilities Council (STFC) Consolidated Grant ST/W001195/1. JDJ acknowledges further funding by the Research Foundation - Flanders (FWO) fellowship 1225625N. This research has made use of NASA's Astrophysics Data System Bibliographic Services.
\end{acknowledgements}

\bibliographystyle{aa}
\bibliography{bibliography.bib}

\newpage
\appendix

\section{\legolas{} convergence tests}
\label{app:lego-conv}

\begin{figure}
    \centering
    \includegraphics[width=\linewidth]{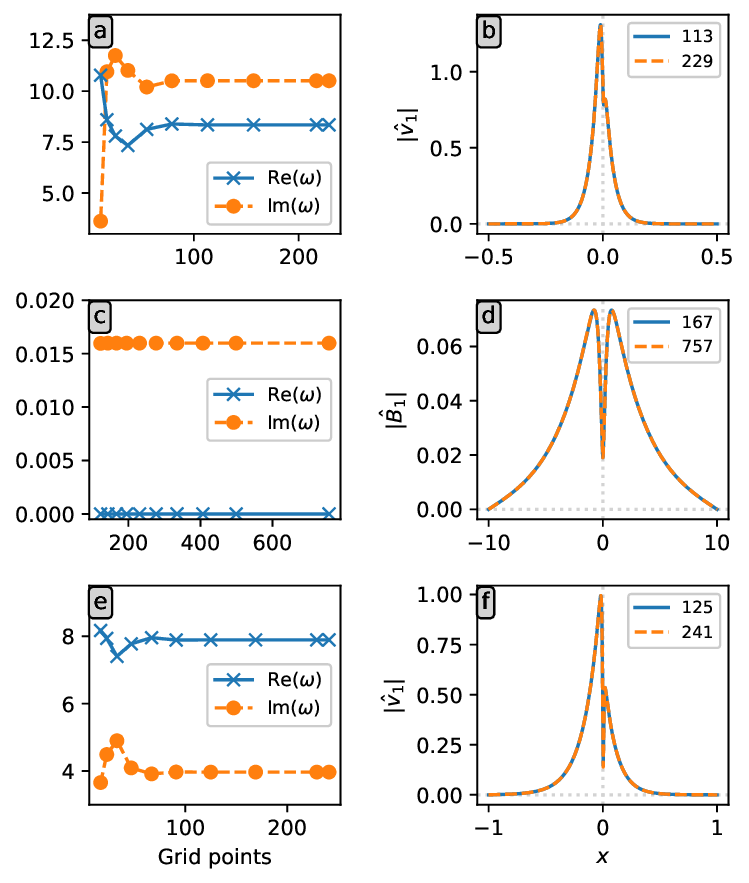}
    \caption{Frequency of the most unstable mode as a function of the \legolas{} grid points and corresponding selected eigenfunction at intermediate and high resolution for (a-b) the flow-sheared fluid interface of Sec. \ref{sec:fluid-khi}, (c-d) the Harris sheet of Sec. \ref{sec:harris}, and (e-f) the flow-sheared plasma interface of Sec. \ref{sec:plasma-khi}.}
    \label{fig:lego-conv}
\end{figure}

To demonstrate that the \legolas{} data used to initialise the simulations have converged, the real and imaginary parts of the most unstable frequency are shown in Fig. \ref{fig:lego-conv}(a,c,e) for each configuration. To obtain the variation in grid points, the parameter $p_3$ of the grid accumulation prescription in \citet{DeJonghe2024}, used throughout this work, was varied between $10^{-3}$ and $10^{-1}$. In all three cases, good convergence is already found at $100$ grid points. Additionally, for each setup, a relevant eigenfunction of the most unstable mode is compared in Fig. \ref{fig:lego-conv}(b,d,f) at intermediate and high resolution, illustrating the low resolution required by the finite elements representation to resolve the perturbation profiles \citep[see e.g.][]{Goedbloed2019}.

\section{Construction and scaling of initial perturbations}
\label{app:scaling}

\begin{table*}
    \centering
    \caption{Overview of all simulations and their initial perturbation.}
    \label{tab:overview}
    \begin{tabularx}{\linewidth}{c|lL}
        Case & Configuration & Initial perturbation \\
        \hline
        1a & Flow-sheared fluid interface & Superposition of different-wavelength KHIs \\
        1b & Flow-sheared fluid interface & Superposition of different-wavelength propagating waves \\
        1c & Flow-sheared fluid interface & Velocity noise \\
        2a & Harris current sheet & Single-wavelength tearing instability \\
        2b & Harris current sheet & Single-wavelength fast wave pair \\
        2c & Harris current sheet & Single-wavelength superposition of a tearing instability and a fast wave pair \\
        2d & Harris current sheet & Analytic $\bfb$ prescription, Eqs. \ref{eq:sen-bx}-\ref{eq:sen-by} \\
        2e & Harris current sheet & Velocity noise \\
        3a & Flow-sheared plasma interface & Superposition of different-wavelength KHIs \\
        3b & Flow-sheared plasma interface & Superposition of different-wavelength propagating waves \\
        3c & Flow-sheared plasma interface & Superposition of all the KHIs and propagating waves in 3a and 3b \\
        3d & Flow-sheared plasma interface & Superposition of the shortest wavelength KHI from 3a and longest wavelength propagating waves from 3b \\
        3e & Flow-sheared plasma interface & Velocity noise
    \end{tabularx}
\end{table*}

To construct an initial perturbation from linear eigenmodes, we distinguish between two approaches. In the first approach, we are only inserting one wavelength, close to the most unstable one, as done in Case 2 (Harris current sheet). In the other approach, we are constructing a superposition with various wavelengths, as done in Cases 1 and 3 (both flow-sheared interfaces) to include the fastest growing mode and break the symmetry between developing vortices.

For a single wavelength, say we are including $m$ eigenmodes calculated with \legolas{}. Then, for a given choice of physical quantity $f$ (e.g. $B_{y}$ for Case 2), the (unscaled) perturbation in any quantity $g \in\{\rho, \bfv, T, \bfb\}$ is constructed as
\begin{equation}
    g^{(1)} = \frac{1}{m} \sum\limits_{i=1}^m \frac{g^{(1)}_i}{\max(f^{(1)}_i)},
\end{equation}
where we denote the perturbation of mode $i$ in $f$ and $g$ as $f^{(1)}_i$ and $g^{(1)}_i$ ($i = 1,\dots,m$), respectively.

Subsequently, to determine the scaling factor $\varepsilon$ for the perturbation, we use the expression for the mean total energy
\begin{equation}
    E = E_\mathrm{kin} + E_\mathrm{int} + E_\mathrm{mag},
\end{equation}
using Eqs. \ref{eq:energies}-\ref{eq:Emag} for the right hand side, expanding each physical quantity as $g = g^{(0)} + \varepsilon g^{(1)}$. Collecting the terms of order $\mathcal{O}(\varepsilon^0)$, we define their sum as the equilibrium energy $E^{(0)}$, and the perturbation energy $E^{(1+)}$ as the sum of all the terms of order $\mathcal{O}(\varepsilon^1)$, $\mathcal{O}(\varepsilon^2)$, and $\mathcal{O}(\varepsilon^3)$. Finally, we equate the energy perturbation $E^{(1+)}$ to a fraction $\alpha$ of the equilibrium energy $E^{(0)}$ and solve
\begin{equation}
    E^{(1+)} = \alpha E^{(0)}
\end{equation}
for $\varepsilon$ to obtain the desired perturbation scaling. Since the solution for $\varepsilon$ is typically not unique, we choose the solution with the smallest absolute value to ensure that all perturbations are as small as possible compared to the equilibrium profiles. An identical scaling procedure is employed for an analytically prescribed perturbation, such as the magnetic field prescription in Case 2d.

The reference simulations were initialised with a random velocity noise perturbation
\begin{equation}\label{eq:noise}
    \bfv_1 = A \left( \cos(2\pi\theta)\,\ex + \sin(2\pi\theta)\,\ey \right)
\end{equation}
with variables $A, \theta$ drawn randomly from a continuous uniform distribution $\mathcal{U}(0,1)$, for each grid cell. For these perturbations, the scaling procedure is modified by utilising the expectation value of the energy $\langle E \rangle$ instead.

When we are working with eigenmodes of various wavelengths, the entire procedure above is first performed for each wavelength separately, including rescaling, after which the different-wavelength solutions are multiplied with a unique phase factor (of size $1$), summed, and rescaled again using the energy scaling procedure.

A summary of all simulations (cases) can be found in Table \ref{tab:overview}, specifying both configuration and initial perturbation.

\section{\amrvac{} convergence tests}
\label{app:amrvac-conv}

\begin{figure}
    \centering
    \includegraphics[width=\linewidth]{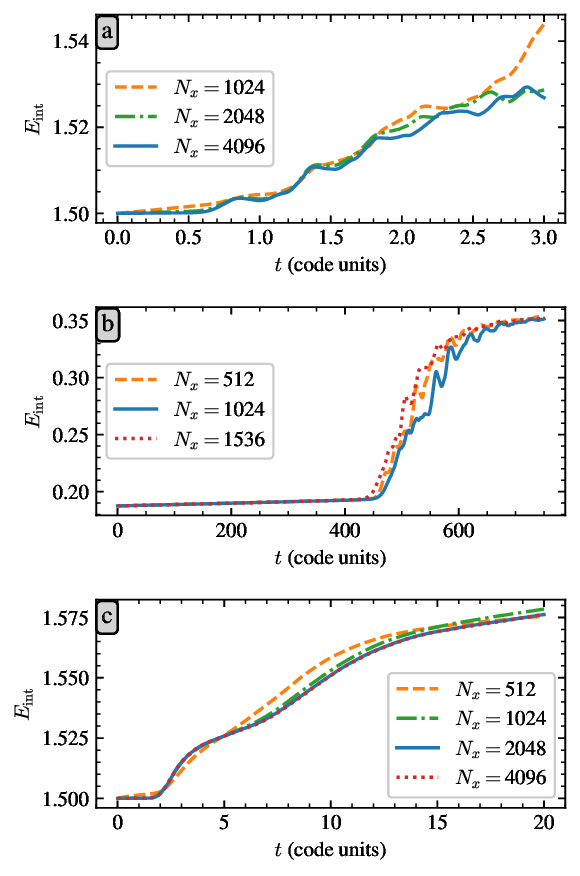}
    \caption{Mean internal energy as a function of time at various resolutions for (a) the flow-sheared fluid interface of Sec. \ref{sec:fluid-khi}, (b) the Harris sheet of Sec. \ref{sec:harris}, and (c) the flow-sheared plasma interface of Sec. \ref{sec:plasma-khi}.}
    \label{fig:amrvac-conv}
\end{figure}

To demonstrate the convergence of the \amrvac{} simulations presented in this work, each configuration's first simulation, initialised with the instability (Case a), was performed at multiple grid resolutions. To evaluate the convergence, we compare the evolution of the spatially averaged internal energy $E_\mathrm{int}$, Eq. \ref{eq:energies}.

For the flow-sheared fluid interface in Sec.~\ref{sec:fluid-khi}, Case 1a was performed using $1024 \times 1056$, $2048 \times 2112$, and $4096 \times 4224$ grid cells. The mean internal energy is shown in Fig. \ref{fig:amrvac-conv}a, with simulations labelled by the number of cells in the $x$-direction, $N_x$. For the Harris sheet in Sec.~\ref{sec:harris}, simulation 2a was repeated using $512 \times 1024$, $1024 \times 2048$, and $1536 \times 3072$ grid cells, with the results shown in Fig. \ref{fig:amrvac-conv}b. For the flow-sheared plasma interface in Sec.~\ref{sec:plasma-khi}, the resolutions are $512 \times 568$, $1024 \times 1136$, $2048 \times 2256$, and $4096 \times 4512$, and the results are shown in Fig. \ref{fig:amrvac-conv}c. In all three cases, the resolution used in the main text for comparing initialisation methods is presented as a solid blue line.

For the flow-sheared interfaces, the energy curves shown in Figs. \ref{fig:amrvac-conv}a and \ref{fig:amrvac-conv}c, the energy curves follow the same trajectory at earlier times, except for the lowest resolution simulation in each case. For the plasma interface (Fig. \ref{fig:amrvac-conv}c), the two highest resolutions coincide perfectly for the whole simulation, proving that our chosen resolution of $2048 \times 2256$ is sufficient. For the fluid interface (Fig. \ref{fig:amrvac-conv}a), the energy curves of the two highest resolution simulations diverge at later times. However, contrary to the plasma interface these simulations develop turbulence, and the chaotic nature of turbulence (butterfly effect) means that very small fluctuations introduced by the difference in discretisations alter the energy evolution at later times. Finally, the Harris sheet simulations in Fig. \ref{fig:amrvac-conv}b show the same behaviour across all resolutions, but differ slightly in onset time (a few percent). Additionally, in the non-linear stage the system reaches a chaotic plasmoid regime, which explains the smaller scale fluctuations in the energy curves.

\end{document}